\documentclass[floatfix,aps,twocolumn,preprintnumbers,amsmath,amssymb,nofootinbib,superscriptaddress,showkeys]{revtex4}

\usepackage{graphicx,color}
\usepackage{amsmath,amssymb,bm}
\usepackage{dcolumn}
\usepackage{comment}
\usepackage{slashed}
\usepackage{ulem}

\newcommand{\bea}{\begin{eqnarray}}
\newcommand{\eea}{\end{eqnarray}}
\newcommand{\be}{\begin{equation}}
\newcommand{\ee}{\end{equation}}
\newcommand{\np}{{\bf p}}

\newcommand{\nh}{{\bf h}}

\newcommand{\nk}{{\bf k}}

\newcommand{\nq}{{\bf q}}

\def\XXint#1#2#3{{\setbox0=\hbox{$#1{#2#3}{\int}$}
     \vcenter{\hbox{$#2#3$}}\kern-.5\wd0}}

\def\1{\'{\i}}

\begin{document}

\title{Meson-exchange currents in quasielastic electron scattering in
  a generalized superscaling approach
}

\author{Paloma R. Casale} \affiliation{Departamento de
  F\'{\i}sica At\'omica, Molecular y Nuclear \\ and Instituto Carlos I
  de F{\'\i}sica Te\'orica y Computacional \\ Universidad de Granada,
  E-18071 Granada, Spain.}

\author{J.E. Amaro}\email{amaro@ugr.es} \affiliation{Departamento de
  F\'{\i}sica At\'omica, Molecular y Nuclear \\ and Instituto Carlos I
  de F{\'\i}sica Te\'orica y Computacional \\ Universidad de Granada,
  E-18071 Granada, Spain.}

\author{M.B. Barbaro}\email{barbaro@to.infn.it} 
\affiliation{
Dipartimento di Fisica, Universit\`a di Torino, 10125 Torino, Italy
}
\affiliation{
 INFN Sezione di Torino, 10125 Torino, Italy
}

\date{\today}

\begin{abstract}
We present a model that incorporates the effect of two-body currents
in quasielastic electron-nucleus scattering  within the
framework of  a consistent superscaling formalism.  This is
achieved by defining an averaged single-nucleon hadronic tensor based
on the 1p1h matrix element of the one-body current plus meson-exchange
currents  (MEC).  The consistent treatment of one- and
two-body currents in our model enables the calculation of exchange
current effects in the kinematical region where the Fermi gas response
is zero, but not the scaling function.  The effect of MEC is
consistently taken into account when extracting the phenomenological
scaling function from electron scattering data.  With this
model, we investigate the effect of MEC on the response functions
taking into account the effective mass of the nucleon, and examine the
consequences it has on the inclusive  $(e,e')$ cross section.
We find that 1p1h MEC deplete the quasielastic transverse
  response, while they not alter significantly the scaling behavior of
  (e,e') data.
\end{abstract}


\maketitle

\section{Introduction}

The pursuit of neutrino oscillation experiments represents a
significant scientific endeavor, encompassing both experimental and
theoretical efforts
\cite{Kami98,Nova16,Agu10,Abe18,Acc14,Ace19}. 
In particular, theoretical nuclear physics has
been propelled to analyze neutrino-induced nuclear reactions in these
experiments
\cite{Alv14,Mos16,Kat17,Ank17,Ben17,Alv18,Ama05,Ama20,Ama21,Ank22}. 
The ultimate goal is to minimize uncertainties arising
from nuclear effects, which are a primary source of systematic errors
when determining the neutrino interactions in
detectors. Simultaneously, there has been a renewed interest in
electron scattering studies \cite{Ama10,Meg16,Ama17,Don23,Don23b}, 
as theoretical models can be calibrated
using (e,e') data and subsequently extended to neutrinos by
incorporating the contribution of the axial current.

At typical energies around 1 GeV in many neutrino experiments, a
significant contribution arises from quasielastic nucleon emission,
which dominates at transferred energies around $\omega=|Q^2|/2m^*_N$,
where $\omega$ is the energy transfer, $Q^2=\omega^2-q^2<0$, and $q$
is the momentum transfer to a nucleon with relativistic effective mass
$m_N^*$ \cite{Ros80,Ser86,Dre89,Weh93,Bar98}.  It is crucial to take into
account that the transferred energies involved in neutrino experiments
necessitate a relativistic treatment of the reaction. This requirement
introduces significant challenges in constructing appropriate models
for these interactions \cite{Giu15,Pan15,Mar16,Gon19,Lov20}.

In this article, we focus on the study of one-particle one hole (1p1h)
transverse and longitudinal responses in the QE peak
\cite{Ama02,Ama03,Ama09}, at intermediate and high momentum transfer,
including the effect of meson exchange currents (MEC) for electron
scattering. The MEC are two-body currents that involve the exchange
of mesons between nucleons and virtual excitation of nuclear
resonances.  This can have a significant impact on the scattering
cross section and on the distribution of energy and momentum
transferred during the interaction.  The emission of two particles
(2p2h), stemming from MEC and short-range correlations, has emerged as
a focal point in studies on lepton-nucleus scattering.

 Extensive
research has been dedicated to understanding its effects on the
cross-section of both electron and neutrino interactions
\cite{Alb84,Van80,Dek94,Mar10,Lal12,Nie13,Cuy17,Pac03,Rui17,Mar21,Mar21b}.
However, it is often overlooked that MEC also contribute to the
emission of a single particle (1p1h), thereby introducing interference
effects with the one-body current. 
Notably, calculations have shown a
reduction in the quasielastic transverse response compared to the
impulse approximation when employing nuclear shell or 
Fermi gas models \cite{Koh81,Alb90,Ama94,Ama03,Meu02,Ama10,Cuy17}.
This reduction is mainly due to the exchange part of the 1p1h 
matrix element of the $\Delta$ current.

In this work, we aim also to incorporate the effect of Meson Exchange
Currents (MEC) consistently into the quasielastic peak within the
framework of the relativistic effective mass Superscaling (SuSAM*)
model \cite{Mar17,Ama18}. This is an extension of SuSA model based on the
approximation of factorizing the nuclear response into a single
nucleon response multiplied by a superscaling function \cite{Don99}. The
phenomenological superscaling function accounts for nuclear structure
and reaction effects, as it is fitted to experimental data.  
The motivation behind
scaling models arises from the observation that inclusive data, when
divided by an appropriate single nucleon prefactor, approximately
scale when plotted against a suitable scaling variable, $\psi$,
extracted from the Relativistic Fermi Gas (RFG) model \cite{Don99}.  
The SuSA
model \cite{Ama05}, along with its improved version SuSAv2, and SuSAM*, has been extensively
utilized to analyze inclusive electron and neutrino scattering data
\cite{Gon14,Meg16,Meg16b,Rui18,Ama21}. 
These efforts represent important strides in understanding and
predicting neutrino-nucleus interactions. By establishing a
phenomenological scaling function that successfully describes (e,e')
data, these models provide a valuable foundation for extrapolating to
neutrino cross-sections.

The SuSAM* model builds upon the SuSA framework but incorporates the
effective mass dependence from the Relativistic Mean Field (RMF)
theory. A notable feature of the RMF model of nuclear matter (such as
the Walecka or $\sigma-\omega$ model \cite{Ser86}) is that it
reproduces the (e,e') cross-section better than the RFG model when an
appropriate value for the effective mass M* is chosen
\cite{Ros80,Weh93,Ama15}.  Motivated by this the SuSAM* model employs
the RMF model's scaling variable, $\psi^*$, and single nucleon
prefactor dependent on the effective mass, with the aim to capture the
essential dynamics associated with the interaction process more
accurately.  This approach capitalizes on the reasonable dynamical
aspects embedded in the RMF model and offers an alternative
description of the scaling behavior observed electron scattering cross
section.  It provides a comprehensive framework that combines the
strengths of the RMF model and the superscaling formalism, leading to
an improved understanding and interpretation of experimental data.

Until now, a unified model that incorporates 1p1h Meson Exchange
Currents in the superscaling function had not been proposed.  This was
primarily due to the violation of scaling properties by MEC, even at
the Fermi gas level \cite{Ama02}.  Additionally, the 1p1h matrix
element of MEC is not easily extrapolated to the $|\psi|>1$ region
outside the range where the Fermi gas response is zero, as nucleons
are constrained by the Fermi momentum.  In this work, we address both
of these challenges in a unified manner by modifying the scaling model
to account for the contribution of MEC within the single nucleon
prefactor. Furthermore, we take the opportunity to enhance the
recently improved superscaling model by eliminating the extrapolation
of single-nucleon responses averaged over the Fermi gas to the region
$|\psi|>1$ \cite{Cas23}.  
Instead of extrapolation, we introduce a new approach where
the single nucleon response is averaged with a smeared momentum
distribution around the Fermi surface. As a result, the averaged
single nucleon responses are well defined for all the values of $\psi$.

In the modified superscaling framework proposed in this work, the
single nucleon response incorporates the contribution of MEC to the
effective one-body current operator. This modification allows us to
define a new prefactor that already includes the effects of MEC,
enabling a novel scaling analysis of the data. Importantly, it should
be noted that the Fermi gas now exhibits exact scaling behavior when
utilizing the new single nucleon response: scaling violations
  associated to the MEC are exactly canceled by the dividing factor
  used to construct the scaling function.  By incorporating these
modifications, we overcome the limitations of previous models and
provide a comprehensive framework that encompasses both MEC and
modified superscaling effects.  By consistently integrating
  1p1h MEC within the SuSAM* model, we aim at refining
 our understanding of the underlying nuclear dynamics in the
quasielastic peak. This comprehensive approach allows us to account
for both the scaling behavior observed in inclusive data and the
contributions from meson exchange currents, leading to a more accurate
and comprehensive description of the reaction.

The article is structured as follows.  In Sect. 2, we introduce the
formalism of quasielastic electron scattering within the framework of
the Relativistic Mean Field (RMF) model of nuclear matter,
incorporating Meson Exchange Currents (MEC).  In Sect. 3, we present
our unified scaling model that incorporates MEC effects. We describe
the modifications made to the conventional scaling approach to account
for the contribution of MEC within the single nucleon prefactor.  In
Sect. 4 we present the results of our calculations and analyses
based on the unified scaling model with MEC. 
Finally in  Sect. 5 we present the conclusions drawn from our study.

\section{Formalism}

\subsection{Response functions}

 We start with the
inclusive electron scattering cross section in plane-wave Born
approximation with one photon-exchange. The exchanged photon
transfers an energy $\omega$ and a momentum $\nq$ to the nucleus.
The initial electron energy is $\epsilon$,  the scattering angle is $\theta$,
and the final electron energy is $\epsilon'=\epsilon-\omega$.
The double-differential cross section is
written in terms of the longitudinal and transverse response
functions, $R_L(q,\omega)$ and $R_T(q,\omega)$,
\begin{equation}  \label{cross}
\frac{d\sigma}{d\Omega d\epsilon'}
= \sigma_{\rm Mott}
\left(v_L R_L(q,\omega) +  v_T  R_T(q,\omega)\right),
\end{equation}
where $\sigma_{\rm Mott}$ is the Mott cross section and 
 $v_{L}$ and $v_{T}$ are the kinematic coefficients defined as
\begin{eqnarray}
  v_{L}&=&\frac{Q^{4}}{q^{4}}\\
  v_{T}&=&\tan\frac{Q^{4}}{q^{4}}-\frac{Q^{2}}{2q^{2}}
\end{eqnarray}
with $Q^{2}=\omega^{2}-q^{2}<0$
the four-momentum transfer.  
The nuclear response functions are the following combinations of the hadronic
tensor
\begin{equation}
  R_L= W^{00},     \kern 1cm
  R_T=W^{11}+W^{22}   .
  \label{eq:responses}
\end{equation}
The inclusive hadronic tensor is constructed from the matrix elements
of the electromagnetic current operator $\hat{J}^{\mu}(\nq)$ 
between the initial and final
hadronic states:
\begin{eqnarray}
W^{\mu\nu}&=& \sum_{f}\overline{\sum_i}
\left\langle f \right|\hat{J}^{\mu}(\nq) |\left. i \right\rangle^{*}
\left\langle f \right|\hat{J}^{\nu}(\nq) |\left. i \right\rangle \nonumber
\\ &\times& \delta(E_f-E_i-\omega),\label{hadronic0}
\end{eqnarray}
where 
 the sum  is performed over the undetected final nuclear states $|f \rangle$ and  the average over 
the initial ground state $|i\rangle$ spin components.

In this work, our approach aims at exploting
the scaling symmetry of
quasielastic data. This scaling symmetry states that the scaling
function, that is, the cross-section divided by  
 an appropriately averaged
single-nucleon cross-section and multiplied by a kinematic factor, only
depends on a single kinematic variable, $\psi$, rather than on the three
variables $(\epsilon, q, \omega)$. The 
scaling function is approximately the same for all nuclei \cite{Ama18}. 
The starting point for the scaling analysis is the relativistic Fermi gas (RFG) model, where
this symmetry holds exactly. In the case of real nuclei, it is only
approximately fulfilled, but it proves to be very useful for analyzing
experimental data and performing calculations and predictions.

\subsection{1p1h hadronic tensor}

In 
independent particle models, the main contribution to the
hadronic tensor in the quasielastic peak comes from the one-particle
one-hole (1p1h) final states. As the transferred energy increases,
there are contributions from two-particle two-hole (2p2h) emission,
the inelastic contribution of pion emission above the pion mass
threshold, and the deep inelastic scattering at higher
energies. Therefore, the hadronic tensor can be generally decomposed
as the sum of the 1p1h contribution and other contributions:
\begin{equation}
W^{\mu\nu}=W^{\mu\nu}_{1p1h}+ W^{\mu\nu}_{2p2h}+ \ldots
\end{equation}
In this work we focus on the 1p1h response which, in the RFG model, reads
\begin{eqnarray}
W^{\mu\nu}_{1p1h}&=& \sum_{ph}
\left\langle
ph^{-1} \right|\hat{J}^{\mu} |\left. F \right\rangle^{*}
\left\langle
ph^{-1} \right|\hat{J}^{\nu} |\left. F \right\rangle 
\nonumber
\\ 
&\times& \delta(E_{p}-E_{h}-\omega)
\theta(p-k_F)\theta(k_F-h)
\label{hadronic}
\end{eqnarray}
where $|p\rangle \equiv |\np s_p t_p\rangle$ and $|h\rangle \equiv
|\nh s_h t_h\rangle$ are plane wave states for particles and holes,
respectively, and $|F\rangle$ is the RFG ground state with all momenta
occupied below the Fermi momentum $k_F$.  The novelty compared to
previous works on scaling is that we start from a current operator
that is a sum of one-body and two-body operators. This approach allows
us to consider the contributions of both the usual electromagnetic
current of the nucleon and the meson-exchange currents (MEC)  to
the 1p1h response:
\begin{equation}
\hat{J}^\mu = 
\hat{J}^\mu_{1} 
+\hat{J}^\mu_{2},
\end{equation}
where $\hat{J}_1$ represents the one-body (OB) electromagnetic current
of the nucleon, while $\hat J_{2}$ is the two-body MEC.  Both currents
can generate non-zero matrix elements for 1p1h excitation. MEC are
two-body operators and they can induce 1p1h excitation due to the
interaction of the hit nucleon with a second nucleon acting as a
spectator.  The many-body matrix elements of these operators are given
by
\begin{equation}
\left\langle ph^{-1} \right|\hat{J}_{1}^{\mu} |\left. F \right\rangle
=
\left\langle p \right|\hat{J}_{1}^{\mu} |\left. h \right\rangle
\end{equation}
for the OB current and
\begin{equation}
\left\langle ph^{-1} \right|\hat{J}_2^{\mu} |\left. F \right\rangle 
=
\sum_{k<k_F}\left[
\left\langle pk \right|\hat{J}_2^{\mu} |\left. hk \right\rangle 
- \left\langle pk \right|\hat{J}_2^{\mu} |\left. kh \right\rangle
\right]
\label{melement}
\end{equation}
for the two-body current, where the sum over 
spectator states $(k)$ is performed over the occupied states 
in the Fermi gas, considering both the direct and exchange matrix elements.
Due to momentum conservation, the  matrix element of the OB current
between plane waves can be written as
 \begin{equation}
 \langle p |\hat{J}_1^{\mu} | h\rangle =
  \frac{(2\pi)^{3}}{V}\delta^{3}(\nq+\nh-\np)
\frac{m_{N}}{\sqrt{E_{p}E_{h}}}
j_1^{\mu}(\np,\nh), 
\label{OBmatrix}
\end{equation}
where $V$ is the volume of the system, $m_{N}$ is the nucleon mass,
$E_p=\sqrt{p^2+m_N^2}$ and
$E_h=\sqrt{h^2+m_N^2}$ are the on-shell energies 
of the nucleons involved
in the process, and
 $j_1^{\mu}(\np,\nh)$ is the OB current (spin-isospin) matrix
\begin{equation}
  j^{\mu}_{1}(\np,\nh)
=\bar{u}(\np)
\left(F_{1}\gamma^{\mu}+i\frac{F_{2}}{2m_{N}}\sigma^{\mu\nu}Q_{\nu}
\right)u(\nh),
\end{equation}
being
$F_{1}$ and $F_{2}$ 
the Dirac and Pauli form factors of the
nucleon.  In the case of the two-body current, the elementary 
matrix element can
be written in a similar form:
\begin{eqnarray}
\langle p'_{1}p'_{2}|\hat{J}_2^{\mu}|p_{1}p_{2}\rangle
&=&
\frac{(2\pi)^{3}}{V^{2}}\delta^{3}(\np_1+\np_2+\nq-\np'_1-\np'_2)
\nonumber\\
&&
\kern -1cm 
{}\times
\frac{m_{N}^{2}}{\sqrt{E'_1E'_2E_1E_2}}
j_2^{\mu}(\np'_1,\np'_2,\np_1,\np_2) .
\label{two-body-matrix}
\end{eqnarray}
Here $j_2^{\mu}(\np'_1,\np'_2,\np_1,\np_2)$ is a spin-isospin matrix
and it depends on the momenta of the two
nucleons in the initial and final state.  The
two-body current contains the sum of the
diagrams shown in Figure 1, including the seagull, pionic, and $\Delta$
isobar currents.  The specific form of the two-body current function
will be given later when we discuss the MEC model.
 By inserting (\ref{two-body-matrix}) into
Eq. (\ref{melement}) we obtain an expression similar to
(\ref{OBmatrix}) that resembles the matrix element of an effective
one-body (OB) current for the MEC:
 \begin{equation}
\left\langle ph^{-1} \right|\hat{J}_2^{\mu} |\left. F \right\rangle 
=
  \frac{(2\pi)^{3}}{V}\delta^{3}(\nq+\nh-\np)
\frac{m_{N}}{\sqrt{E_{p}E_{h}}}
j_2^{\mu}(\np,\nh).
\end{equation}
Here the effective OB current generated by the MEC involves a sum over
the spectator nucleons and is defined  by
\begin{eqnarray}
j_2^{\mu}(\np,\nh) 
&\equiv &
\nonumber\\
&&
\kern -1cm
\sum_{k<k_F}
\frac{m_{N}}{VE_k}
\left[ j_2^{\mu}(\np,\nk,\nh,\nk)-j_2^{\mu}(\np,\nk,\nk,\nh)\right] .
\label{effectiveOB}
\end{eqnarray}
Note that in the  thermodynamic limit 
$V \rightarrow \infty$ the above sum will be transformed into 
 an integral over the momenta occupied in the Fermi gas:
\begin{equation}
\frac{1}{V}\sum_{k<k_F}
\rightarrow 
\sum_{s_kt_k}\int \frac{d^3k}{(2\pi)^3} \theta(k_F-k) .
\end{equation}
Finally, we can write the transition matrix element of the total
current between the ground state and the 1p1h state as
 \begin{equation}
\left\langle ph^{-1} \right|\hat{J}^{\mu} |\left. F \right\rangle 
=
  \frac{(2\pi)^{3}}{V}\delta^{3}(\nq+\nh-\np)
\frac{m_{N}}{\sqrt{E_{p}E_{h}}}
j^{\mu}(\np,\nh), 
\label{total}
\end{equation}
where the effective total current for the 1p1h excitation 
includes contributions from both the one-body current and MEC:
\begin{equation}
j^{\mu}(\np,\nh)= j_1^{\mu}(\np,\nh)+ j_2^{\mu}(\np,\nh). 
\end{equation}
By inserting (\ref{total})  into Eq. (\ref{hadronic})  and taking the thermodynamic limit, we obtain the following expression for the hadronic tensor:
\begin{eqnarray}
W^{\mu\nu}&=&
\frac{V}{(2\pi)^{3}}\int{d^3h\delta(E_{p}-E_{h}-\omega)
\frac{m_{N}^{2}}{E_{p}E_{h}}}2w^{\mu\nu}(\np,\nh) \nonumber \\
&\times&
\theta(p-k_{F})\theta(k_{F}-h),\label{integralw}
\end{eqnarray}  
where $\np=\nh+\nq$ by momentum conservation after integration over $\np$.
The  function $w^{\mu\nu}$ is the effective single-nucleon hadronic tensor 
in the transition 
\begin{eqnarray}
w^{\mu\nu}(\np,\nh)=\frac{1}{2}\sum_{s_ps_h}
j^\mu(\np,\nh)^*j^\nu(\np,\nh) .
\end{eqnarray}
 In this
equation, we did not include the sum over isospin $t_p=t_h$. 
Therefore, $w^{\mu\nu}$ refers to the
tensor of either proton or neutron emission, and the total tensor
would be the sum of the two contributions.
Note that the effective single-nucleon tensor $w^{\mu\nu}$ includes the
contribution of MEC, thus encompassing an interference between the
one-body and two-body currents.
Indeed, the relevant diagonal components of the effective
single-nucleon hadronic tensor for the longitudinal and transverse
responses \eqref{eq:responses} can be expanded as 
\begin{eqnarray}
w^{\mu\mu}(\np,\nh)
&=& \frac{1}{2}\sum_{s_ps_h} |j^\mu_1+j^\mu_2|^2
\nonumber\\
&=& 
\frac{1}{2}\sum |j^\mu_1|^2
+\mbox{Re}\sum (j^\mu_1)^*j^\mu_2
+\frac{1}{2}\sum |j^\mu_2|^2
\nonumber\\
&\equiv&
  w^{\mu\mu}_1+ w^{\mu\mu}_{12}+w^{\mu\mu}_2
\end{eqnarray}
where \(w^{\mu\mu}_1\)
is the tensor corresponding to the one-body current, \(w^{\mu\mu}_{12}\)
represents the interference between the one-body and two-body
currents, and \(w^{\mu\mu}_2\) corresponds to the contribution of the two-body
current alone.
The one-body part is the leading contribution in the quasielastic peak,
 while the dominant
contribution of the MEC corresponds to the interference with the
one-body current   \cite{Ama94,Ama03},
 being the pure contribution of the two-body current  generally smaller.

\subsection{Responses in the relativistic mean field approach}

Going beyond the Relativistic Fermi Gas (RFG) model, the Relativistic
Mean Field (RMF) approach for nuclear matter allows for the inclusion
of dynamic relativistic effects. The simplest approximation in this
framework is to introduce constant mean scalar and vector potentials
with which the nucleons interact \cite{Ros80,Ser86,Weh93,Bar98}. 
 The scalar potential is attractive,
while the vector potential is repulsive. The single-particle wave
functions still exhibit plane-wave behavior with momentum \(p\) in
nuclear matter, but with an on-shell energy given by 
\begin{equation}
E = \sqrt{m_N^ {*2}  + p^2},
\end{equation}
where \(m_N^*\) is the relativistic effective mass of the nucleon, defined as
\begin{equation}
m_N^*=m_N-g_s\phi_0 = M^*m_N.
\end{equation}
Here  $\phi_0$
is the scalar potential energy of the RMF and $g_s$ the corresponding coupling constant
\cite{Ser86}, and $M^*=0.8$ for $^{12}$C, the nucleus considered in
this work \cite{Mar21}.  To account for the interaction with the
vector potential, a positive energy term needs to be added to the
on-shell energy. Therefore, the total energy of the nucleon can be
expressed as:
\begin{equation}  \label{energyrmf}
E_{RMF}=E+E_v.
\end{equation}
In this work we use the value $E_v=141$ MeV, obtained in
Ref. \cite{Mar21} for $^{12}$C.  Note that in observables that only
depend on the energy differences between initial and final particles,
the vector energy cancels out, and only the on-shell energy
appears. This cancellation happens, as we will see, in the response associated
to the one-body current. However, in the case of the two-body current,
the vector energy needs to be taken into account in the $\Delta$ current,
as we will see in the next section. 

In the present RMF approach of nuclear matter, the evaluation of the
hadronic tensor is done similarly to the RFG, with the difference that
the spinors \(u(p)\) now correspond to the solutions of the Dirac
equation with the relativistic effective mass \(m_N^*\). 
From Eq. (19) the 1p1h nuclear
response functions are then given by
 \begin{eqnarray}
R_K(q,\omega)
&=& 
 \frac{V}{(2 \pi)^3}
\int
d^3h
\frac{(m^*_N)^2}{E_pE_h}  2w_K(\np,\nh)  
\nonumber\\
&& 
\kern -1cm 
\times \theta(p-k_F)
\theta(k_F-h) 
\delta(E_p-E_h-\omega),
\label{rmf}
\end{eqnarray}
where $E_p, E_h$ are the on-shell energies with effective mass $m_N^*$,
and $w_K$ are the single-nucleon responses for the 1p1h excitation
\begin{eqnarray} \label{snresponses}
w_L =  w^{00},
\kern 1cm
w_T  =  w^{11}+w^{22}.
\end{eqnarray}
The effective single-nucleon tensor $w^{\mu\nu}$ is constructed as in
Eq. (20), but the current is obtained from matrix elements using
spinors with the relativistic effective mass \(m_N^*\) instead of the
normal nucleon mass. This prescription is also followed when
evaluating the 1p1h matrix elements of the MEC (as discussed in the
next section).

To compute the integral (\ref{rmf}), we change to the variables
$(E_h, E_p, \phi)$, using
$h^2 dh d\cos\theta= (E_hE_p/q)dE_hdE_p$.  Then the  integral over $E_p$ can be { performed
using
 the Dirac delta.  This fixes the angle $\theta_h$ 
between $\nq$ and $\nh$ 
\begin{equation} \label{angulo}
\cos\theta_h= \frac{2E_h\omega+Q^2}{2hq}. 
\end{equation}
The integration over the angle $\phi$ gives
$2\pi$ by symmetry of the responses when $\nq$ is on the $z$-axis
\cite{Ama20}. The result is an integral over the initial nucleon
energy
 \begin{equation}
R_K(q,\omega)
= 
 \frac{V}{(2 \pi)^3}
 \frac{2\pi m_N^{*3}}{q}
\int_{\epsilon_0}^{\infty}d\epsilon\, n(\epsilon)\, 2w_K(\epsilon,q,\omega),
\label{respuesta}
\end{equation}
where we have defined the adimensional energies $\epsilon=E_h/m^*_N$ and
$\epsilon_F=E_F/m_N^*$.  Moreover we have introduced the energy
distribution of the Fermi gas $n(\epsilon)=
\theta(\epsilon_F-\epsilon)$.  The lower limit of the integral
(\ref{respuesta}), \(\epsilon_0\), represents the minimum energy that
an on-shell nucleon can have when it absorbs energy \(\omega\) and
momentum \(q\) \cite{Ama20}
\begin{equation}
\epsilon_0={\rm Max}
\left\{ 
       \kappa\sqrt{1+\frac{1}{\tau}}-\lambda, \epsilon_F-2\lambda
\right\},
\end{equation}
where we have defined the dimensionless variables 
\begin{equation}
\lambda  = \frac{\omega}{2m_N^*},
\kern 1cm 
\kappa   =  \frac{q}{2m_N^*}, 
\kern 1cm
\tau  =  \kappa^2-\lambda^2. 
\end{equation}
 From Eq. (28) the nucleons that contribute to the response function
\(R{_K}(q,\omega)\) are those with energy ranging from \(\epsilon_0\) to
\(\epsilon_F\).  For fixed values of $\phi, q, \omega$, the integral
over energy $\epsilon$ in Eq. (\ref{respuesta}) corresponds to
integrating the single nucleon response over a path in the momentum
space of the hole $\nh$, weighted with the momentum distribution. The
angle between $\nh$ and $\nq$ for each energy is given by Eq. (27).
The minimum momentum $h_0$ correspond to the minimum energy
$\epsilon_0$. Indeed, for a specific value of \(\omega\), the lower
limit of the integral becomes \(h = 0\) or \(\epsilon_0 = 1\), which
corresponds to the center of the quasielastic peak. Using Eq. (29), it
is straightforward to verify that this point corresponds to \(\lambda
= \tau\) in the regime without Pauli blocking.

\subsection{Scaling}

Scaling is based on the approximated factorization of an averaged
single-nucleon response from the nuclear cross section. This
factorization is exact in the RMF model with the OB current.  
In previous works,
analytical expressions were obtained from the RFG and RMF models by
explicit integration of the one-body responses, Eq. (\ref{respuesta}).
 However in
this case, it is not possible to perform the integration
(\ref{respuesta}) analytically because now $w_K$ includes also the
matrix elements of the two-body operator.  Nevertheless, we can still
define averaged single-nucleon responses as
\begin{eqnarray}\label{average}
\overline{w}_K(q,\omega) = \frac{\int^{\infty}_{\epsilon_0} d \epsilon
  \, n (\epsilon) w_K(\epsilon,q,\omega)}{\int_{\epsilon_0}^{\infty}
  d\epsilon \, n (\epsilon)}
\end{eqnarray}
and we  can rewrite Eq. (\ref{respuesta}) in the form
 \begin{equation}
R_K(q,\omega)
=
 \frac{V}{(2 \pi)^3}
 \frac{2\pi m_N^{*3}}{q}  2\overline{w}_K(q,\omega)
\int_{\epsilon_0}^{\infty}d\epsilon\, n(\epsilon).
\label{respuesta2}
\end{equation}
The averaged single-nucleon responses, $\overline{w}_K(q,\omega)$,
include the combined effect of both the OB current and the MEC in all
the 1p1h excitations compatible with given values of $(q,\omega)$.
Eq. (\ref{respuesta2}) shows that in the RMF model (or the RFG model for
effective mass $M^*=1$) 
the nuclear responses factorize as the product of the averaged single-nucleon 
response (including MEC) and the scaling function.
In fact a superscaling function can be defined as
\begin{equation} \label{scaling0}
 f^*(\psi^*)=
\frac{3}{4}\frac{1}{\epsilon_F-1}
\int_{\epsilon_0}^{\infty} n (\epsilon) d\epsilon,
\end{equation}
where $\epsilon_F-1$ is the kinetic Fermi energy in units of
$m_N^*$  
and  the $\psi^*$-scaling variable is related to the minimum nucleon energy,
$\epsilon_0$, as 
\begin{equation} \label{psi}
\psi^* = \sqrt{\frac{\epsilon_0-1}{\epsilon_F-1}} {\rm sgn} (\lambda-\tau).
\end{equation}
The scaling variable, $\psi^*$, is negative (positive) for
$\lambda<\tau$ ($\lambda>\tau$).
In the RMF the scaling function is easily evaluated from
 Eq. (\ref{scaling0}), giving
\begin{equation} \label{scalingRFG}
 f^*(\psi^*)=\frac{3}{4}(1-\psi^*{}^2)\theta(1-\psi^*{}^2).
\end{equation}
Note that  the scaling function of nuclear matter is zero for
$\epsilon_0 > \epsilon_F$, and this is equivalent to $|\psi^*| > 1$.
This is a consequence of the maximum momentum $k_F$ for the nucleons
in nuclear matter, which implies that $\epsilon_0< \epsilon_F$.

Using $V/(2\pi)^3= N/(\frac83 \pi k_F^3)$ for nuclear matter
we can write the response functions (\ref{respuesta2}) as 
\begin{equation}
R_K(q,\omega) = 
\frac{\epsilon_F-1}{m_N^* \eta_F^3 \kappa}
 (Z \overline{w}^p_K(q,\omega)+N \overline{w}^n_K(q,\omega))
f^*(\psi^*),
\label{susam} 
\end{equation}
where we have added the contribution of $Z$ protons and $N$ neutrons
to the response functions, and $\eta_F=k_F/m_N^*$.

\subsection{SuSAM*}

The expression given by Eq (\ref{susam}) for the response function is formally
the same as the response in the RMF,
the only difference being that the averaged single-nucleon response
now includes the contribution of MEC to
the 1p1h excitation. This equation, valid for the RMF, serves as the
starting point for performing the superscaling analysis with
relativistic effective mass (SuSAM*) using electron scattering data, extending
the formula to the region \(\epsilon_0 > \epsilon_F\) or \(|\psi^*| >
1\). We will follow the procedure suggested by Casale {\it et al.}
\cite{Cas23}.

In the Fermi gas, it is not possible to extend the averaging formula
for \(\epsilon_0 > \epsilon_F\) because the momentum distribution is
zero and the denominator in (\ref{average}) vanishes. 
Therefore, what we do is slightly
modify the Fermi gas distribution by allowing a smeared Fermi surface,
so that the distribution is not exactly zero above \(k_F\), allowing
for the averaging procedure. By substituting the Fermi distribution
with a distribution that is not significantly different from the
original one for \(h < k_F\), the average of the single-nucleon
response will not change significantly in the Fermi gas region $|\psi^*|<1$.

By this method, the extension of the single-nucleon average is done
smoothly and continuously to the region \(|\psi^*| > 1\), with the
added meaning that, in this way, we are taking into account, at least
partially, the high-momentum distribution. This is because it is
primarily the nucleons with momenta greater than \(k_F\) that
contribute to this region.  A possible distribution that can be used
to extend the averaging procedure is the Fermi distribution:
\begin{equation}
n(h)=\frac{a}{1+{\rm e}^{(h-k_F)/b}} .
\label{eq:Fermi}
\end{equation}
Using this distribution, the integrals in the numerator and
denominator of Eq (\ref{average}) extend to infinity and are
well-defined for \(\epsilon_0 > \epsilon_F\) or \(|\psi^*| > 1\). An
appropriate value for the smearing parameter is \(b = 50\) MeV/c, used
in ref. \cite{Cas23}, where the averaged single-nucleon responses were
evaluated for the one-body current, and it was found to yield
practically the same results as the analytically calculated responses
in the strict Fermi gas region.
The averaged responses were also found to be
very similar to the traditionally extrapolated responses outside  this region.
This proposed method provides a simple approach
that allows for the definition of generalized scaling, including the
MEC, consistently, and also takes into account that the nucleons are
not limited by a maximum Fermi momentum.

Several approaches exist to obtain a phenomenological scaling
function. Different methods are based on different assumptions for the
scaling function or the single-nucleon response, but all are
ultimately adjusted to experimental data. The original SuSA model,
based on the RFG, was fitted to the scaling data of the longitudinal
response, to obtain a longitudinal scaling function, $f_L$, while in
the extended SuSA-v2 approach, 
 the RMF model for finite nuclei was used to obtain
a transverse scaling function, $f_T$ .  The SuSAM* model, based on the 
 nuclear matter
RMF with effective mass, directly fitted the quasielastic data of the
cross section after discarding the non-scaling data points, to obtain
a single phenomenological scaling function valid for both the L and T
channels \cite{Ama17}.

In the generalized SuSAM* model proposed here, we will follow the same
procedure as described in references [42, 43]. First, we subtract the
calculated inclusive cross section for two-particle emission in the
RMF with a relativistic MEC model from the (e,e') data. This
subtraction aims to partially remove the contribution of 2p2h
processes present in the data, in order to isolate the purely
quasielastic data as much as possible. Next, we will scale each
 residual
data point by dividing it by the contribution of the single
nucleon to the cross section, as given by Eq. (36),
\begin{equation} \label{fexp}
f_{exp}^* =
\frac
{\displaystyle 
\left(\frac{d\sigma}{d\Omega d\omega}\right)_{exp}
  -\left(\frac{d\sigma}{d\Omega d\omega}\right)_{2p2h}
}
{\sigma_M ( v_L r_L + v_T  r_T) },
\end{equation}
where the single nucleon cross section includes the averaged
single-nucleon responses including MEC
\begin{equation} \label{rsn}
r_K = \frac{\epsilon_F-1}{m_N^* \eta_F^3 \kappa} 
(Z \overline{w}^p_K(q,\omega)+N \overline{w}^n_K(q,\omega)).
\end{equation}
In the results section, we will proceed with the scaling analysis for
the obtained $f_{exp}^*$ data, by a plot as a function of $\psi^*$,
calculated using Eq. (\ref{psi}). This analysis includes a selection
process to identify the data points that are most likely to be
quasielastic (which exhibit approximate scaling behavior) and discarding
the remaining data points (mainly non-scaling inelastic
processes). Finally, we will fit a phenomenological scaling function
to the surviving data points, aiming to describe the global scaling
behavior of the quasielastic region.

\begin{figure}[t]
\centering
\includegraphics[width=7.5cm,bb=120 460 495 700]{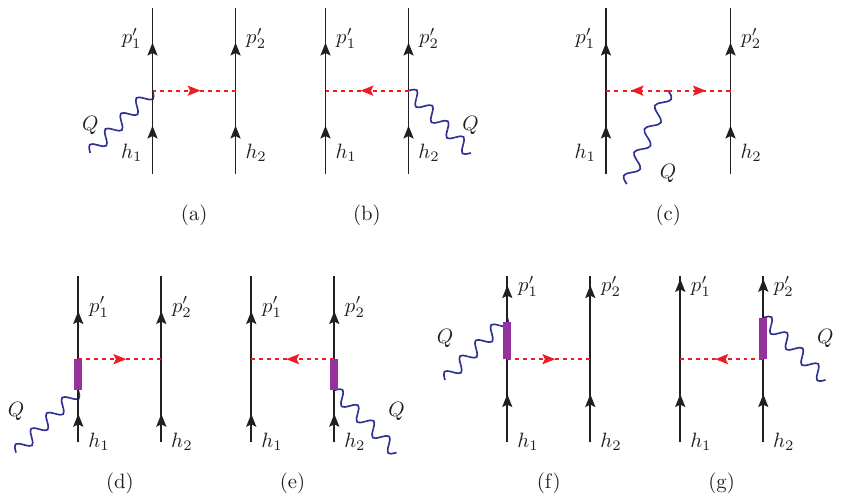}
\caption{Feynman diagrams for the 2p2h MEC model used in 
this work.}\label{feynman}
\end{figure}

\subsection{Meson-exchange currents}

In this work, we use the relativistic meson exchange currents
(MEC) model described in Ref. \cite{Rui17}. The Feynman diagrams shown
in Fig. 1 illustrate the different components of the MEC
model. Diagrams (a) and (b) correspond to the seagull current, diagram
(c) represents the pion-in-flight current, and diagrams (d,e) and
(f,g) depict the forward- and backward- $\Delta(1232)$ currents,
respectively. The specific treatment of the $\Delta$ current is
model-dependent, and various versions exist with possible corrections
to the off-shell relativistic interaction of the $\Delta$. Other widely
used models for MEC include those described in Refs. \cite{Pac03,Ama03,Pas95}.

While these different models may exhibit slight variations and
corrections to the $\Delta$ off-shell interaction, they generally yield
similar results for the dominant transverse response at the
quasielastic peak. In particular, in the results section, we will
compare our findings with the model presented in Refs. \cite{Pas95,Ama03},
which we previously employed to assess the impact of MEC on the 1p1h
response.

In our model the MEC functions defined in Eq. (\ref{two-body-matrix})
correspond to the sum of diagrams of Fig. 1
\begin{equation}
j_2^\mu(\np'_1,\np'_2,\np_1,\np_2) 
=  j^{\mu}_{sea}+  j^{\mu}_{\pi}+  j^{\mu}_{\Delta},
\end{equation}
where the $\Delta$ current is the sum of forward and backward terms
\begin{equation}
 j^{\mu}_{\Delta}= j^{\mu}_{\Delta F}+ j^{\mu}_{\Delta B}.
\end{equation}
These functions are defined by  
\begin{widetext}
\begin{eqnarray} \label{2p2h}
  j^{\mu}_{sea}
&=&
[I_{z}]_{t'_1t'_2,t_1t_2}
\frac{f^{2}}{m_{\pi}^{2}}
V^{s'_{1}s_{1}}_{\pi NN}(p'_{1},p_{1})
F_{\pi NN}(k_{1}^{2})
\bar{u}_{s'_2}(p_{2}^{'})F_{1}^{V}\gamma^{5}\gamma^{\mu}u_{s_2}(p_{2}) 
+ (1 \leftrightarrow 2)
\\
  j^{\mu}_{\pi}
&=&[I_{z}]_{t'_1t'_2,t_1t_2}
\frac{f^{2}}{m_{\pi}^{2}}F_{1}^{V}
V^{s'_{1}s_{1}}_{\pi NN}(p'_{1},p_{1})
V^{s'_{2}s_{2}}_{\pi NN}(p'_{2},p_{2})(k_{1}^{\mu}-k_{2}^{\mu}) 
\\  
  j^{\mu}_{\Delta F}
&=&
[U_{z}^{F}]_{t'_1t'_2,t_1t_2}
\frac{ff^{*}}{m_{\pi}^{2}}
V^{s'_{2}s_{2}}_{\pi NN}(p'_{2},p_{2})F_{\pi N \Delta}(k_{2}^{2})
\bar{u}_{s'_1}(p'_{1})k_{2}^{\alpha}G_{\alpha\beta}(p_{1}+Q)
\Gamma^{\beta\mu}(Q)u_{s_1}(p_{1}) 
+ (1 \leftrightarrow 2)
\label{delta1}
\\
  j^{\mu}_{\Delta B}
&=&
[U_{z}^{B}]_{t'_1t'_2,t_1t_2}
\frac{f^{2}f^{*}}{m_{\pi}^{2}}
V^{s'_{2}s_{2}}_{\pi NN}(p'_{2},p_{2})
F_{\pi N \Delta}(k_{2}^{2})
\bar{u}_{s'_1}(p'_{1})k_{2}^{\beta}
\hat{\Gamma}^{\mu\alpha}(Q)G_{\alpha\beta}(p'_{1}-Q)u_{s_1}(p_{1}) 
+ (1 \leftrightarrow 2)  
\label{delta2}
\end{eqnarray}
\end{widetext}
We will evaluate these matrix elements in the framework of the RMF
model, where the spinors \(u(p)\) are the solutions of the Dirac equation
with relativistic effective mass \(m_N^*\).  The four-vectors
$k_{i}^{\mu}=p'_{i}{}^{\mu}-p_{i}^{\mu}$ with $i=1,2$ are the momenta
transferred to the nucleons 1,2.  We have defined the following
function that includes the $\pi NN$ vertex, a form factor, and the
pion propagator
\begin{equation}
  V^{s'_{1}s_{1}}_{\pi NN}(p'_{1},p_{1})
= F_{\pi NN}(k_{1}^{2})
\frac{ \bar{u}(p'_{1})\gamma^{5}\slashed{k}_{1}u(p_{1})}{k_1^2-m_{\pi}^2}.
\end{equation}
We apply strong form factors at the pion absorption/emission 
vertices given by  \cite{Alb84,Som78} 
\begin{equation}
  F_{\pi NN}(k)=   F_{\pi N\Delta}(k)
=\frac{\Lambda^{2}-m_{\pi}^{2}}{\Lambda^{2}-k^{2}}.
\label{pinnff}
\end{equation}
The charge structure of the MEC involves the isospin matrix element
of the operators
\begin{eqnarray}
  I_{z} &=& i[\tau(1) \times \tau(2)]_z,
\\  
 U^{F}_z 
&=&
 \sqrt{\frac{3}{2}}\sum_{i=1}^{3}{(T_{i}T_{z}^{\dagger}) \otimes \tau_{i}},
\\ 
  U^{B}_z 
&=& \sqrt{\frac{3}{2}}\sum_{i}^{3}{(T_{z}T_{i}^{\dagger}) \otimes \tau_{i}},  
\end{eqnarray}
where we denote by $T^{\dagger}_i$ the Cartesian coordinates of the 
$\frac{1}{2} \rightarrow \frac{3}{2}$
transition isospin operator, defined by its matrix elements \cite{Eri88}  
\begin{equation}
\textstyle
\langle \frac32 t_\Delta | T^\dagger_\mu | \frac12 t \rangle
= \langle \frac12 t 1 \mu | \frac32 t_\Delta \rangle
\end{equation}  
$T^\dagger_\mu$ 
 being the spherical components of the vector
$\vec{T}^\dagger$.  With the aid of the expression
$T_{i}T^{\dagger}_{j}=(2/3)\delta_{ij}-\frac{i}{3}\tau_{i}\tau_{j}$
and making the summation, we can rewrite the isospin operators in the
forward and backward $\Delta$ current as
\begin{align}
  U^{F}_z&=\sqrt{\frac{3}{2}}\left(\frac{2}{3}\tau_{z}^{(2)}-\frac{i}{3}\left[\tau^{(1)}\times\tau^{(2)}\right]_{z}\right)\\
 U^{B}_z&=\sqrt{\frac{3}{2}}\left(\frac{2}{3}\tau_{z}^{(2)}+\frac{i}{3}\left[\tau^{(1)}\times\tau^{(2)}\right]_{z}\right).    
\end{align}
The $\gamma N\rightarrow \Delta$ transition vertex in the
forward $\Delta$ current is defined as \cite{Lle72,Her07}
\begin{equation}
  \Gamma^{\beta\mu}(Q)=
\frac{C_3^V}{m_N}
(g^{\beta\mu}\slashed{Q}-Q^{\beta}\gamma^{\mu})\gamma_5
\end{equation}
while for the backward $\Delta$ current
\begin{equation}
 \hat{ \Gamma}^{\mu\alpha}(Q)=\gamma^{0}[\Gamma^{\alpha\mu}(-Q)]^{\dagger}\gamma^{0}.
\end{equation}
In this vertex we have neglected contributions of order $O(1/m_N^2)$.
Note that the $\Gamma^{\beta\mu}$ operator is a spin matrix 
and depends on the vector form factor $C_3^V$. 
In this paper, we use the vector form factor in $\Delta$ current from
Refs. \cite{Nie13,Her07}:
\begin{equation}
  C_{3}^{V}(Q^{2})
=\frac{2.13}{(1-\frac{Q^{2}}{M_{V}^{2}})^{2}}
\frac{1}{1-\frac{Q^{2}}{4M_{V}^{2}}} .
\end{equation}
Various alternative
approximations to the propagator have been proposed \cite{Qui17}.
 However, in the
case of the quasielastic peak, the typical kinematics are  of
the order
of 1 GeV, and these issues are not expected to be relevant. They are
overshadowed by other more significant nuclear effects that dominate
in this energy regime.
Here we use the $\Delta$ propagator commonly used for the spin-3/2 field
\begin{equation}
  G_{\alpha\beta}(P)=
\frac{{\cal P}_{\alpha\beta}(P)}{
P^{2}-M_{\Delta}^{2}+iM_{\Delta}\Gamma(P^{2})+\frac{\Gamma(P^{2})^{2}}{4}}
\end{equation} 
where $M_{\Delta}$ and $\Gamma$ are the $\Delta$ mass and width
respectively. The projector ${\cal P}_{\alpha\beta}(P)$ over spin-3/2 on-shell
particles is given by
\begin{eqnarray}
{\cal  P}_{\alpha\beta}(P)
&=&
-(\slashed{P}+M_{\Delta})
\nonumber\\
&&
\kern -1cm
 \times
\left[
g_{\alpha\beta}-\frac{\gamma_{\alpha}\gamma_{\beta}}{3}-\frac{2P_{\alpha}P_{\beta}}{3M_{\Delta}^{2}}+\frac{P_{\alpha}\gamma_{\beta}-P_{\beta}\gamma_{\alpha}}{3M_{\Delta}}
\right] .
\end{eqnarray}  
Finally, the 
$\Delta$ width $\Gamma(P^{2})$ is given by
\begin{equation}
  \Gamma(P^{2})=\Gamma_{0}\frac{m_{\Delta}}{\sqrt{P^{2}}}
\left(\frac{p_{\pi}}{p^{res}_{\pi}}\right)^{3}.
\label{width}
\end{equation}
In the above equation, $p_{\pi}$ is the momentum of the final pion
resulting from the $\Delta$ decay an $p^{res}_{\pi}$ is its value at
resonance ($P^2=m_\Delta^2$), and $\Gamma_{0}=120$ MeV is the width at
rest.
The width (\ref{width}) corresponds to the $\Delta$ in vacuum,
and it is expected to be slightly different in the medium depending on
the kinematics.
 One could investigate the dependence of
the results on the choice of the width. However, in this work, we do
not delve into this issue because, as we will see, the effect of the
MEC on the 1p1h response is generally small, and corrections due to
fine-tuning of the model are unlikely to substantially alter the
results.

\begin{figure}[t]
  \centering
\includegraphics[width=8cm,bb=110 460 500 690]{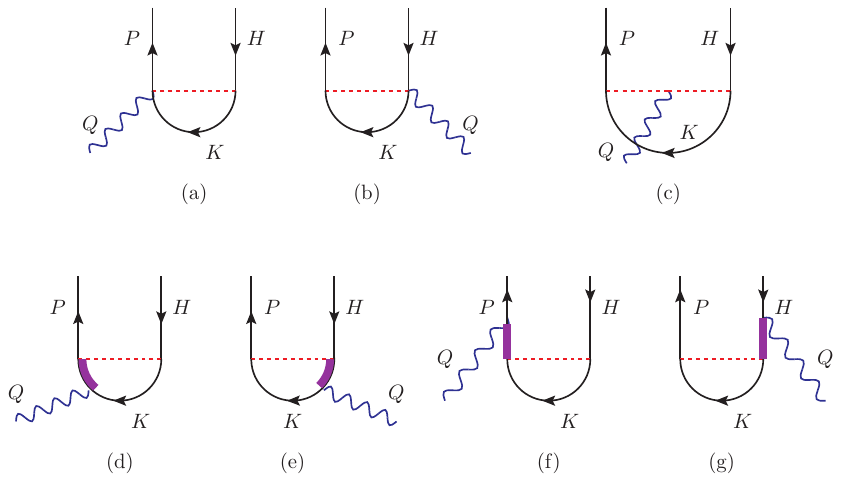}
\caption{Diagrams for the 1p1h MEC matrix elements}
\label{feymnan2}
\end{figure}

In the relativistic mean field description used in this work, we
consider that the $\Delta$ is also interacting with scalar and vector
fields, acquiring an effective mass and vector energy. To treat this
case, we make the following substitutions in the $\Delta$ propagator 
for the $\Delta$ mass and momentum
\cite{Weh93,Kim96}: 
\begin{equation} 
M_\Delta \rightarrow M_\Delta^*,
\kern 1cm
P^{*\mu} = P^\mu - \delta_{\mu0}E_v^\Delta.
\end{equation}
   We use the value $M^*_\Delta=1042$ MeV, taken from
\cite{Mar21b}, and the universal vector coupling $E_v^\Delta=E_v$.

With the MEC current defined in Eqs. (24-27), the effective one-body
current \(j_2(\np,\nh)\) is generated by summing over  the spin, isospin and momentum of the spectator nucleon, as in  Eq. (15). First, it
can be observed that due to the sum over isospin \(t_k\), the direct
term is zero (see Ref. \cite{Ama03} for details). Therefore, the
many-body diagrams that contribute to the 1p1h MEC are those shown in
Figure 2. Furthermore, it can be verified that diagrams e and f are
also zero.
Therefore, only diagrams a, b, c,
and d survive and contribute to the 1p1h MEC matrix elements.

\section{Results}

In this section, we present results for the effects of MEC on the
1p1h response functions using several models: the relativistic Fermi
gas, the relativistic mean field, and the generalized SuSAM* model. By
employing these different models, we 
take into account relativistic kinematics and we can analyze  the impact of including the relativistic effective
mass of the nucleon and the $\Delta$ resonance appearing in the MEC.
The scaling analysis described in the previous Section will allow us to study the influence of MEC on
the generalized scaling function also in the region $|\psi^*|>1$ where the 
RFG and
RMF responses are 
zero. Moreover,  we can investigate how the inclusion of MEC
affects the scaling function
and compare it with the predictions of
the RFG and RMF models.

\begin{figure}
  \centering
\includegraphics[width=6cm,bb=147 475 380 798]{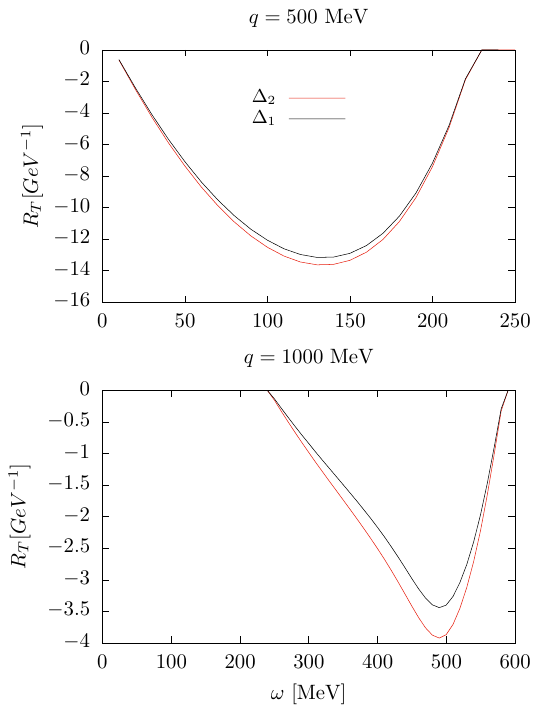}
\caption{Interference OB-MEC in the transverse response of $^{40}$Ca
  for two values of the momentum transfer, with $k_F=237$ MeV/c.  In
  the graph, the curve labeled $\Delta_1$ corresponds to using the $\Delta$
  current of the present work in RFG. The curve $\Delta_2$ corresponds to the
  calculation from reference \cite{Ama03}.
}
\label{figpas}
\end{figure}

\begin{figure}[t]
  \centering
\includegraphics[width=8cm,bb=100 350 450 800]{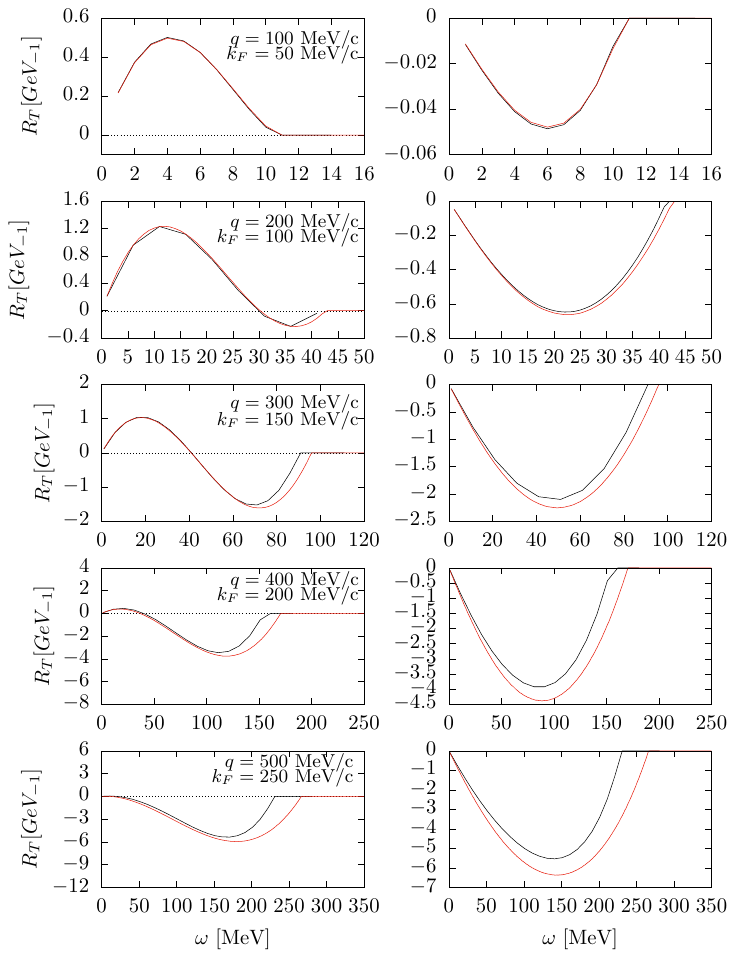}
\caption{Comparison  
between relativistic and non relativistic  MEC transverse responses in
  $^{12}$C.  Black lines: RFG.  Red lines: non-relativistic
  Fermi gas. Left panels show the interference OB-$\pi$, 
  and left panels the interference OB-$\Delta$. In these calculations the
  strong form factors in the pion vertices are set to one.}
\label{fig_scaling}
\end{figure}

Unless stated otherwise, we present the results for $^{12}$C with a
Fermi momentum of $k_F = 225$ MeV/c. We use
an effective mass of $M^* = 0.8$, following the same choice of
parameters as in reference \cite{Mar21,Mar21b}.  The calculation of
1p1h responses involves evaluating the 1p1h matrix element of the MEC,
as given by Eq (15). This requires performing a numerical
three-dimensional integration to account for the momentum
dependence. Subsequently, a one-dimensional integration is carried out
to calculate the averaged single-nucleon responses, as described in Eq
(31).

First, since this work is an extension of the MEC model from
Ref. \cite{Ama03} to the superscaling formalism, we will compare with
the OB-MEC interference responses presented in \cite{Ama03} within the
framework of the RFG. It should be noted that
in \cite{Ama03} a different version of the $\Delta$ current was used.
The $\Delta$ current was obtained from the \(\gamma N \Delta\) Lagrangian
proposed by Pascalutsa \cite{Pas95}
\begin{equation} \label{pascalutsa}
{\cal L}_{\gamma N \Delta} = ie\frac{G_1}{2m_N}
\overline{\psi}^\alpha \Theta_{\alpha\mu}\gamma_\nu\gamma_5T_3^\dagger 
N F^{\nu\mu} + \mbox{h.c.},
\end{equation}
plus $O(1/m_N^2)$ terms that give negligible contribution 
 in  the quasielastic energy region.  The tensor
$\Theta_{\alpha\mu}$ may contain an off-shell parameter and another
arbitrary parameter related to the contact invariance of the
Lagrangian.
In this work we use the simplest form
\begin{equation}
\Theta_{\alpha\mu}=g_{\alpha\mu}-\frac14\gamma_\alpha\gamma_\mu.
\end{equation}
The coupling constant $G_1$  was determined in \cite{Pas95} 
by fitting Compton
scattering on the nucleon. However, there is a detail
that needs to be clarified:  
the isospin operator used by
 Pascalutsa
is normalized differently from the
standard convention. That is, \(T_i^{\text{Pascalutsa}} =
\sqrt{\frac{3}{2}} T_i\), where \(T_i\) is the operator used in our
calculation. This means that if we use the standard \(T_i\) in
the Lagrangian (\ref{pascalutsa}), it should be multiplied by
\(\sqrt{\frac{3}{2}}\). This is equivalent to multiplying Pascalutsa's
coupling constant $G_1=4.2$ by the factor \(\sqrt{\frac{3}{2}}\). In reference
\cite{Ama03}, this detail went unnoticed, and the $\sqrt{3/2}$ factor was not
included in the calculations.

Using the Lagrangian given by Eq. (\ref{pascalutsa}), the following $\Delta$ current is obtained:
\begin{widetext}
\begin{eqnarray}
  j^{\mu}_{\Delta F}
&=&
[(T_iT_3^\dagger)\otimes\tau_i]_{t'_1t'_2,t_1t_2}
\frac{ff^{*}}{m_{\pi}^{2}} F_\Delta(Q^2)
V^{s'_{2}s_{2}}_{\pi NN}(p'_{2},p_{2})F_{\pi N \Delta}(k_{2}^{2})
\nonumber\\
&&
\bar{u}_{s'_1}(p'_{1})k_{2}^{\alpha}
\left[
\Theta^{\alpha\beta}G_{\beta\rho}(p_{1}+Q)
\frac{G_1}{2m_N}
[ \Theta^{\rho\mu}\gamma^\nu-\Theta^{\rho\nu}\gamma^\mu]\gamma_5Q_\nu
\right]u_{s_1}(p_{1}) 
+ (1 \leftrightarrow 2)
\end{eqnarray}
\end{widetext}
and a similar expression for the $\Delta$ backward current. 
This current was used in Ref. \cite{Ama03} to compute the  OB-MEC interference
with the following form factor
\begin{equation}
  F_\Delta(Q^{2})=
G^{p}_{E}(Q^{2})\left(1-\frac{Q^{2}}{3.5(GeV/c)^{2}}\right)^{-1/2}  
\end{equation}  
where $G_{E}^{p}$ is the electric form factor of the proton.

\begin{figure*}[t]
\centering
\includegraphics[width=8cm,bb=110 270 460 770]{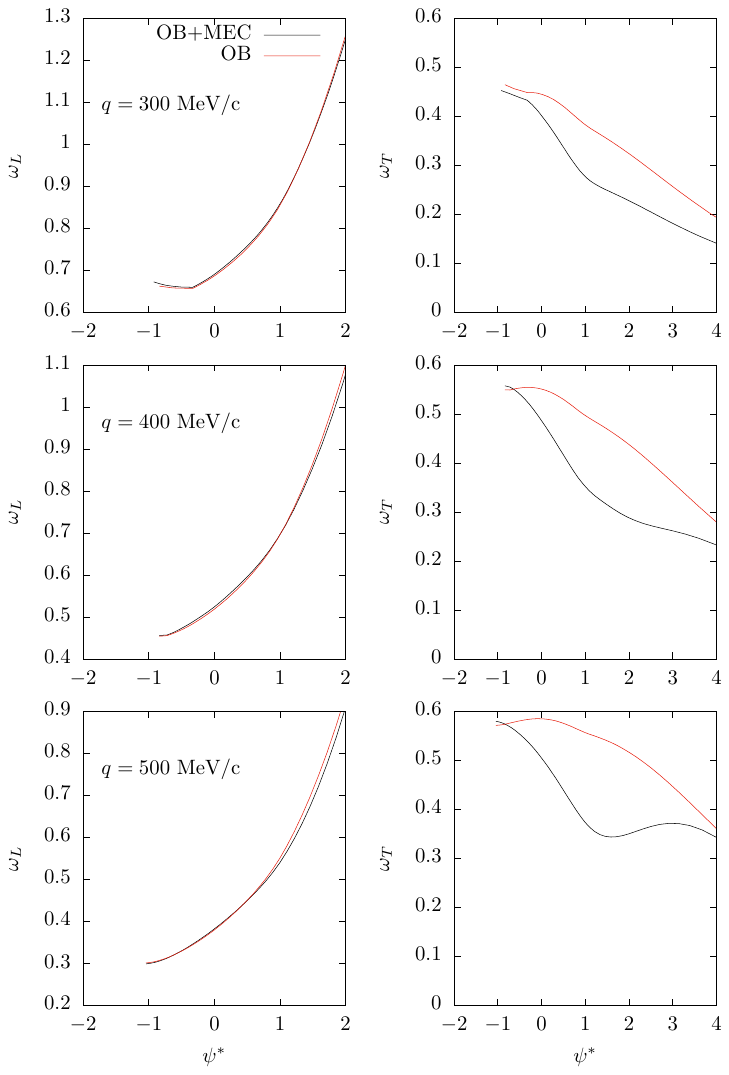}
\kern 4mm
\includegraphics[width=8cm,bb=110 270 460 770]{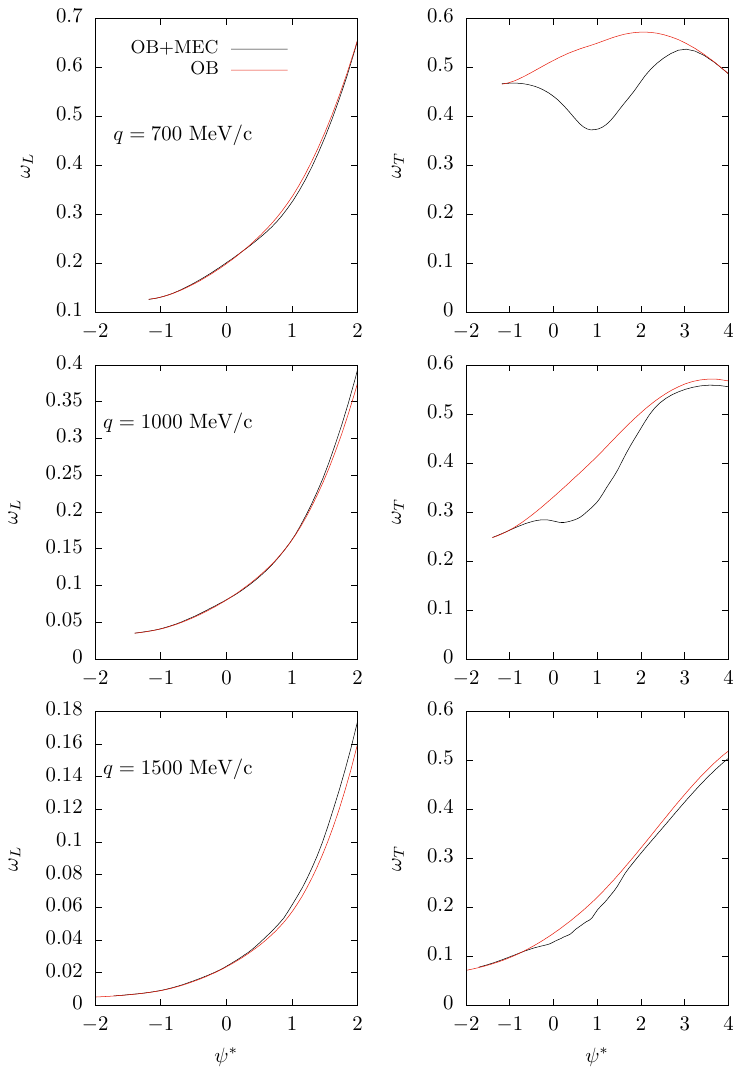}
  \caption{Averaged single nucleon responses computed with and without
    MEC, for several values of the momentum transfer as a function of
    the scaling variable $\psi^*$.  }
  \label{fsingle}
\end{figure*}

In Figure \ref{figpas}, we present the interference between the OB and
$\Delta$ currents in the transverse response of $^{40}$Ca. We compare
our results with the model of reference \cite{Ama03} in RFG, where the
Lagrangian of Pascalutsa was used. The results of \cite{Ama03} have
been corrected with the factor of \(\sqrt{\frac{3}{2}}\) mentioned
earlier.  
For \(q = 500\) MeV/c, there is
little difference between the two models. However, for \(q = 1\)
GeV/c, the difference becomes more noticeable.

The results of Fig. \ref{figpas} show that the $\Delta$ current
model used in this work does not differ significantly from the model
in reference \cite{Ama03}, providing similar results. The small
differences observed can be attributed to the different form factor
and coupling constants,
and can be understood as a model dependence in these results.
From here on, all the results refer to the $\Delta$ 
current model described in the equations (\ref{delta1},\ref{delta2}).

It is expected that any relativistic model should reproduce the
results of the well-established non-relativistic model for small
values of energy and momentum in the non-relativistic limit 
\cite{Hok73}.  As a
check in this regard, in Fig. 4 we compare the present model with the
non-relativistic Fermi gas model from ref. \cite{Ama94}. The non
relativistic $\Delta$ current used is taken from \cite{Ama03}.  To
perform this comparison the same form factors and coupling constants
are used in the relativistic and non relativistic models.  To take
this limit in Fig. 4, we follow the procedure as follows: \(q\) is
small and \(k_F = q/2\).  We show the comparison between
the two models for various values of \(q\) ranging from 100 to 500
MeV/c. In the left panels, we present 
the contribution of the transverse response stemming  from the 
 interference 
{ OB-$\pi$} between the pure pionic MEC (diagrams a-c in Fig.2) 
and in the right panels we show the OB-$\Delta$
interference (diagrams d-g in Fig.2) for the same values of \(q\). 
As expected, we observe
that for \(q = 100\) MeV/c,  the relativistic and non-relativistic
models practically coincide, demonstrating the consistency between the
two models in the non-relativistic limit.

In Fig.4 one can also observe that for low values of \(q\) the
dominant contributions to the MEC are from the seagull and
pion-in-flight diagrams, with the seagull diagram playing a
particularly important role. These diagrams contribute positively to
the MEC, enhancing the overall response. On the other hand, the
contribution from the $\Delta$ resonance is 
 negative. As \(q\) increases, the
influence of the $\Delta$ resonance becomes more significant, and it
starts to dominate the MEC contribution for \(q\) values around 400
MeV/c.

Before performing the scaling analysis, we examine the averaged
single-nucleon responses that will be used to scale the data. In
Figure \ref{fsingle}, we display the longitudinal and transverse
single-nucleon responses for various values of \(q\) as a function of
the scaling variable. The calculated responses are shown separately
for the OB current and the total responses including the MEC and
taking into account the sum of protons and neutrons.  The total
response, which we have defined in equation (\ref{susam}), comes from
the product of the single nucleon with the phenomenological scaling
function obtained from the $(e,e')$ data as shown below.  We have used
a Fermi distribution, Eq.\eqref{eq:Fermi}, with a smearing parameter
$b=50 MeV/c$, although the single nucleon responses do not depend much
on this specific value.  It is observed that the effect of the MEC is
negligible in the longitudinal response, as the curves for the OB
current and total response overlap. However, in the transverse
response, the effect of the MEC becomes appreciable, resulting in a
reduction of the \(w_T\) response compared to the OB current. This
reduction can be attributed to the interference between the one-body
and two-body currents, which leads to a modified transverse
response. The comparison between the OB current and the total response
including the MEC provides insights into the contributions of the MEC
to the single-nucleon responses and sets the stage for the subsequent
scaling analysis.

\begin{figure}[t]
  \centering
\includegraphics[width=7cm,bb=150 270 400 790]{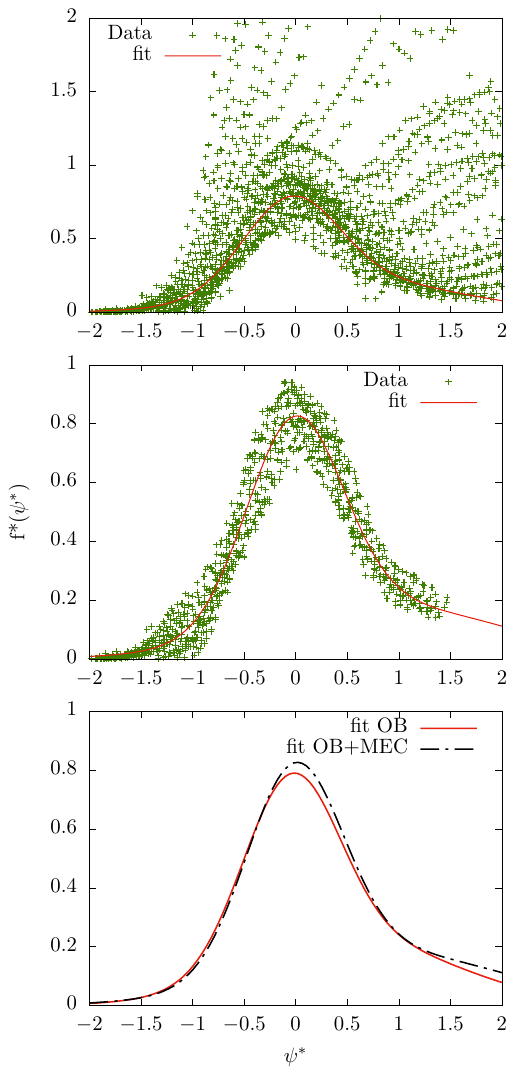}
\caption{Scaling analysis of ${}^{12}$C data including MEC and
  relativistic effective mass $M^{*}=0.8$. The Fermi momentum is
  $k_F=225$ MeV/c.  In the top panel, we show the data points after
  scaling, representing the overall distribution. In the middle panel,
  we display the selected data points, which have been chosen after
  eliminating those that do not exhibit clear scaling behavior.  In
  the bottom panel, we present the phenomenological scaling function,
  which has been fitted to the selected data points, compared to the
  scaling function obtained in a similar analysis without MEC.
  Experimental data are taken from Refs. \cite{archive,archive2}.  }
\label{fig-scaling}
\end{figure}

Note that the center of the quasielastic peak corresponds to
\(\psi^*=0\), where the energy and momentum can be transferred to a
nucleon at rest.  We see that MEC have a larger impact in the region
$\psi^*> 0$, that is, the right-hand side of the peak, corresponding
to higher energy transfers.

In Figure \ref{fig-scaling}, we present the scaling analysis of the
$^{12}$C data. In the top panel, the experimental data, $f^*_{exp}$,
are plotted against $\psi^*$ in the interval
$-2<\psi^*<2$. Experimental data are from
Refs. \cite{archive,archive2} and cover a wide electron energy range,
from 160 MeV up to 5.8 GeV. We observe a
significant dispersion of many data points, indicating a wide range of
inelastic scattering events. However, we also notice that a portion of
the data points cluster together and collapse into a thick band. These
data points can be considered 
as associated to quasielastic (1p1h) events. To select
these quasielastic data, we apply a density criterion. For each point,
we count the number of points, \(n\), within a neighborhood of radius
\(r=0.1\), and eliminate the point if \(n\) is less than 25.  Points
that have been disregarded are likely to correspond to
inelastic excitations and low energy processes that violate scaling
and cannot be considered within quasielastic processes. We observe
that the remaining selected points, about half of the total, shown in
the middle panel of Fig \ref{fig-scaling}, form a distinct thick
band. These points represent the ones that best describe the
quasielastic region and approximately exhibit scaling behavior. The
red curve represents the phenomenological quasielastic function
$f^*(\psi^*)$, that provides the best fit to the selected data using a
sum of two Gaussian functions:
\begin{equation}
  f^{*}(\psi^{*})=
a_{3}e^{-(\psi^{*}-a_{1})^{2}/(2a_{2}^{2})}
+b_{3}e^{-(\psi^{*}-b_{1})^{2}/(2b_{2}^{2})}.
\end{equation}
The parameters found are shown in table \ref{table:1}.

\begin{table}[t]
\begin{ruledtabular}
\begin{tabular}{cccccc}
$a_{1}$ & $a_{2}$ & $a_{3}$ & $b_{1}$ & $b_{2}$ & $b_{3}$ \\
\hline
$-0.01015$ & 0.46499 & 0.69118  & 0.86952 & 1.16236  & 0.17921 \\ 
\end{tabular}
\end{ruledtabular}
\caption{Table of fitted parameters of the scaling function.}
\label{table:1}
\end{table}

In the bottom panel of Fig. \ref{fig-scaling}
 we compare the scaling function obtained in our
analysis with the scaling function obtained without including the MEC
contributions. When including the MEC, the scaling function appears
slightly higher since the single-nucleon response with MEC is slightly
smaller than without them. However, both analyses provide a similarly
acceptable description of the data. This suggests that while the MEC
do have an impact on the scaling behavior, their effect is relatively
small and does not significantly alter the overall scaling pattern
observed in the data.

\begin{figure*}[t]
\centering
\includegraphics[width=13cm,bb=10 550 520 800]{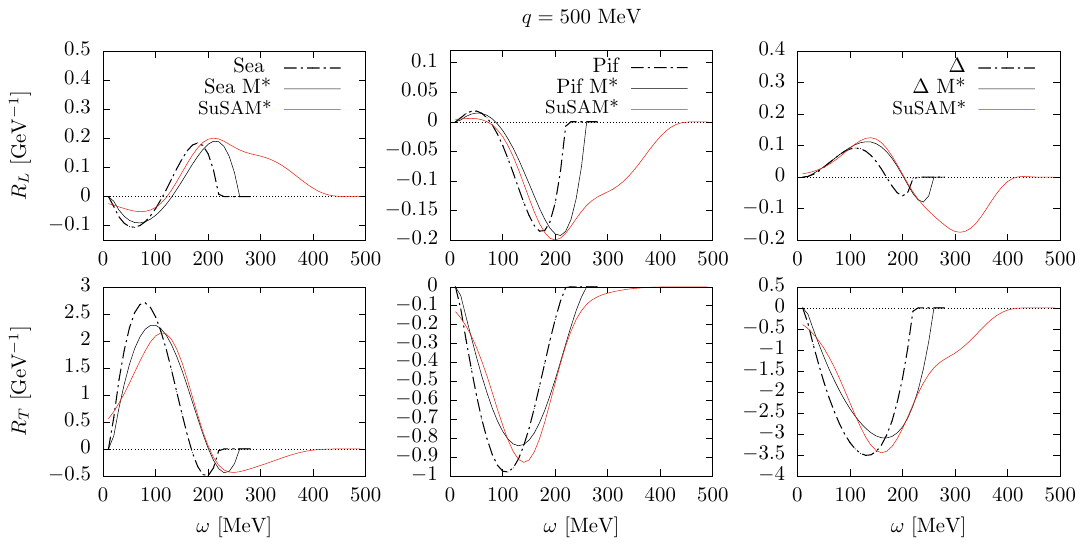}
\caption{Interference OB-MEC responses separated in seagull,
  pion-in-flight, and $\Delta$ contributions for $^{12}$C and $q=500$ MeV/c. In
  each panel we compare the results of RFG (with $M^*=1$,  dot-dash), with the
  RMF (with $M^*=0.8$) and the SuSAM* model. }
\label{fig1}
\end{figure*}

\begin{figure*}[t]
\centering
\includegraphics[width=13cm,bb=10 550 520 800]{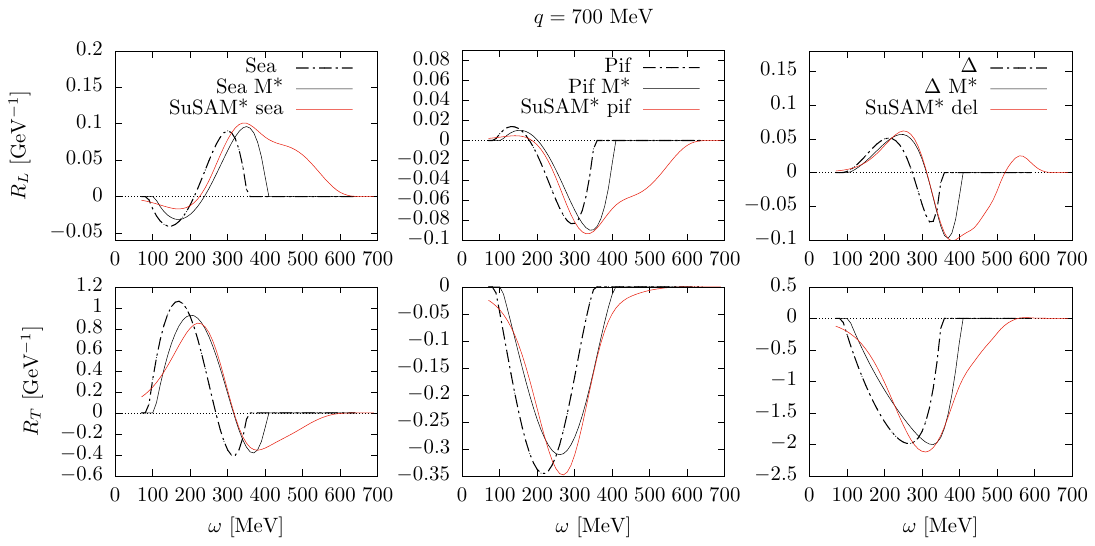}
\caption{
The same as Fig. 7 for $q=700$ MeV/c.
}
\label{fig1}
\end{figure*}

\begin{figure*}[t]
\centering
\includegraphics[width=13cm,bb=10 550 520 800]{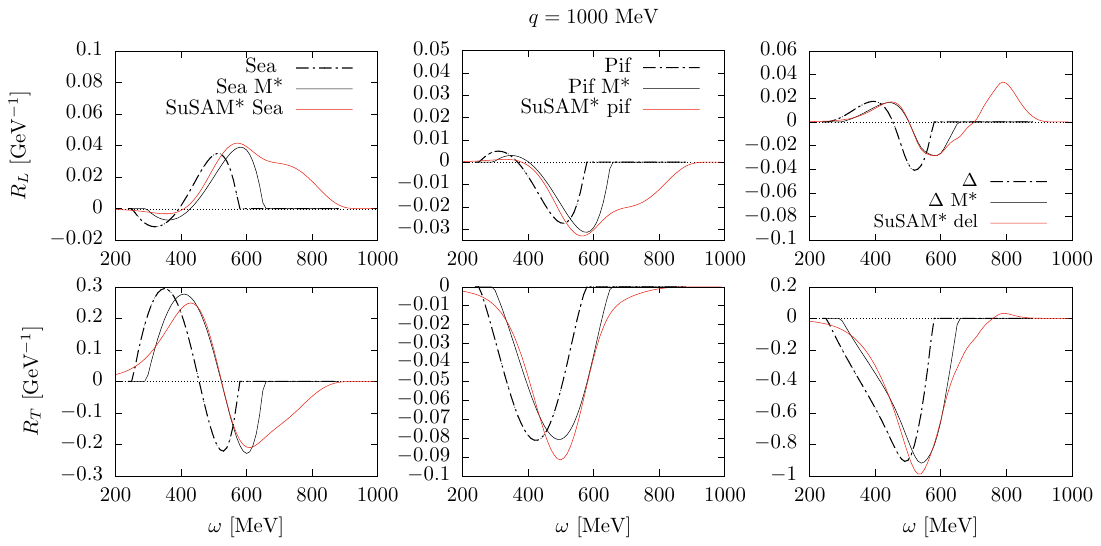}
\caption{The same as Fig. 7 for $q=1000$ MeV/c.
}
\label{fig2}
\end{figure*}

\begin{figure*}[t]
\centering
\includegraphics[width=13cm,bb=10 550 520 800]{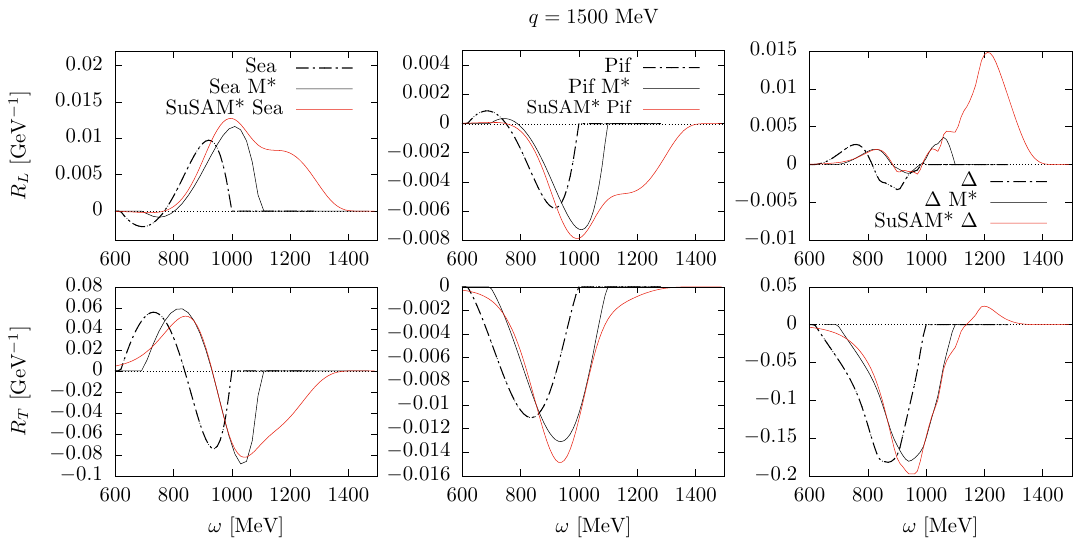}
\caption{The same as Fig. 7 for $q=1500$ MeV/c.
}
\label{fig3}
\end{figure*}

Now that we have obtained the phenomenological scaling function
through the scaling analysis, we can utilize this function to
calculate the response functions of the model beyond the RMF. By
 multiplying
the scaling function by
the averaged single nucleon
responses, as stated in Eq (36), we can extend our calculations to
different kinematic regimes and explore the behavior of the responses
beyond the relativistic mean field description. This allows us to
investigate the influence of various factors, such as the MEC and
relativistic effects, on the response functions and cross sections.

In Figures 7-10, we present the interferences of the OB-MEC in the
response functions for different values of \(q\) (500, 700, 1000, and
1500 MeV/c). We separate the interferences into OB-seagull, OB-pionic,
and OB-$\Delta$ contributions for both the longitudinal and transverse
responses as functions of \(\omega\). Each panel displays three curves
corresponding to the free RFG (with effective mass \(M^*=1\)), the RMF
(with effective mass \(M^*=0.8\)), and the present SuSAM* model.
These figures allow us to analyze the relative contributions of the
different OB-MEC interferences in the response functions at various
kinematic regimes. By comparing the results obtained from the RFG,
RMF, and SuSAM* models, we can observe the effects of including
the relativistic interaction through 
the effective mass and the scaling function on the interferences. 

First is observed that the introduction of the effective mass $M^{*}=0.8$
shifts the responses to the right, towards higher energy values.
 The effective mass takes into account the binding of
the nucleon in the nucleus, which causes the quasielastic peak to
approximately coincide with the maximum of the experimental cross
section. In the RFG, this is traditionally taken into account by
subtracting a binding energy of approximately 20 MeV from \(\omega\)
to account for the average separation energy of the nucleons. In the
RMF, this is automatically included by considering the effective mass
of the nucleon, \(M^*=0.8\), which was adjusted for \(^{12}\)C
precisely to achieve this effect.

In the transition from the RMF to the SuSAM* model, we replace the
scaling function of the RFG with the phenomenological scaling function
that we have adjusted. This new scaling function extends beyond the
region of $-1 < \psi^* < 1$, where the RFG scaling function is
zero. As a result, we observe in figures 7-11 that the interferences
acquire a tail towards high energies, similar to the behavior of the
scaling function.

The tail effect is more pronounced in the longitudinal responses
because the single-nucleon longitudinal response, as shown in Figure
5, increases with $\omega$. This amplifies the tail when multiplied by
the scaling function. However, it is important to note that the
contribution of the MEC to the longitudinal response is relatively
small compared to the dominant transverse response. Therefore, while
the tail effect is observed in the longitudinal responses, its impact
on the cross section is not as significant as in the transverse
channel, if not negligible.

In the dominant transverse response, the seagull contribution from the
MEC is positive,
leading to an enhancement of  the response, while the pionic
and $\Delta$ contributions are negative, causing a reduction in the
overall response when including the MEC. This is in line with
pioneering calculations by 
Kohno and Otsuka \cite{Koh81} and by  Alberico
{\it et al.} \cite{Alb90} in the non-relativistic Fermi gas.  Also in
shell model calculations, similar results have been obtained
\cite{Ama94}, showing that the MEC contributions also lead to a tail
and extension of the response functions to higher values of $\omega$,
as in the SuSAM* approach. 
It is worth noting that the relative
importance of these contributions can depend on the momentum transfer
$q$ and the energy transfer $\omega$. For the values considered in
Figures 7-11, the $\Delta$ current is found to be the dominant
contribution, leading to a net negative effect from the MEC.

The observation in Fig. 10 of a sign change and a small bump in the
OB-$\Delta$ transverse response for high values of $\omega$ is indeed
interesting. The change of sign is already observed for $q$=1 GeV/c in
Fig. 9.  This connects with the findings in reference \cite{Ama10},
where a pronounced bump and sign change were reported in a
semi-relativistic shell model calculation based on the Dirac equation
with a relativistic energy-dependent potential.  
In the present calculation the bump is observed but it is very small
compared to the results of Ref, \cite{Ama10}.  It is important to note
that, in the present work, the fully relativistic SuSAM* approach is
employed, which takes into account the dynamical properties of both
nucleons and the $\Delta$, as well as the scaling function. This
differs from the approach in reference \cite{Ama10}, where a static
propagator for the $\Delta$ was used.  To definitively clarify the
difference with the present results, a fully relativistic calculation
in finite nuclei, considering the dynamical properties of the $\Delta$
would be necessary.

\begin{figure}[t]
\includegraphics[width=6cm,bb=170 370 380 800]{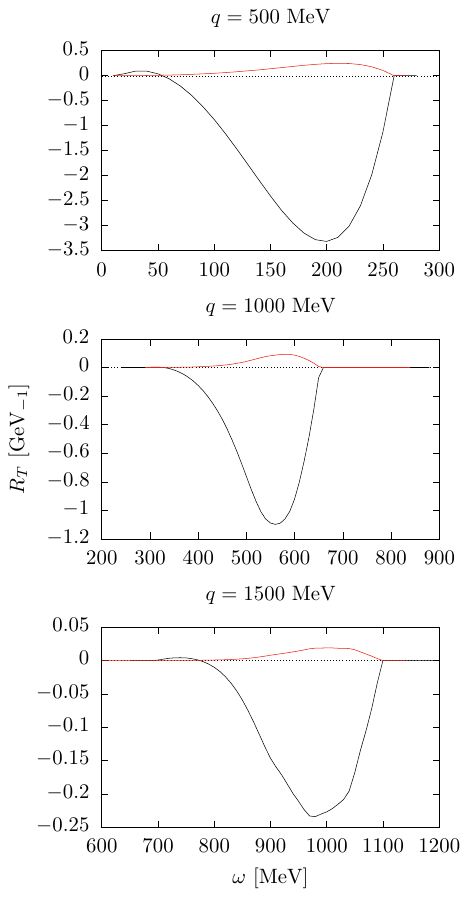}
  \caption{Comparison of OB-MEC interference in the transverse
    response (black lines) with the pure MEC transverse response (red
    lines) for several values of $q$ in the RMF model.}
 \label{fig4}
\end{figure}

\begin{figure}[t]
  \includegraphics[width=6cm,bb=170 370 380 800]{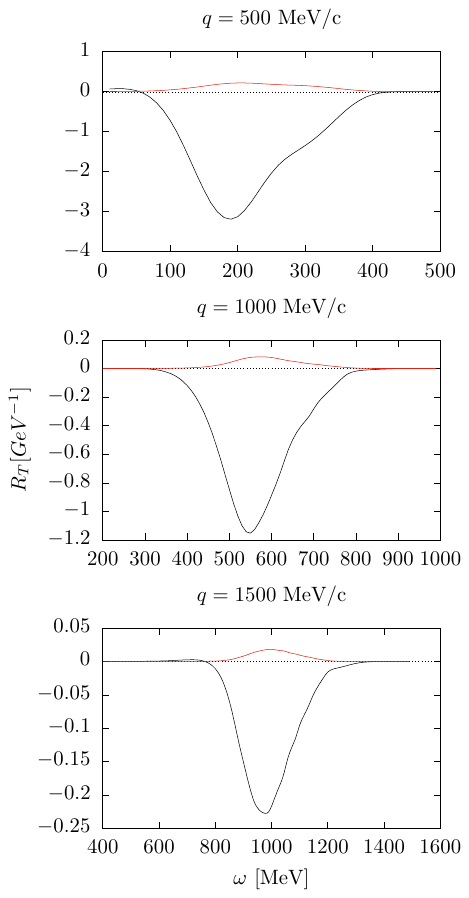}
  \caption{The same as Fig. 11 in the SuSAM* model.}
\end{figure}

The comparison of the OB-MEC interference with the MEC contribution
alone (represented by $w_{12}^{\mu\nu}$ and $w_2^{\mu\nu}$,
respectively in Eq, (21)) in the transverse response is shown in
Figs. 11 and 12. We observe that the MEC contribution alone represents
a small and almost negligible contribution to the transverse
response. This justifies the previous calculations that focused only
on the OB-MEC interference (e.g., the semi-analytical calculations in
references \cite{Ama94a,Ama94} for the non-relativistic Fermi gas), as
it provides an excellent approximation.  This observation holds true
for both the RMF model in Fig. 11 and the SuSAM* model in Fig. 12. It
highlights the fact that the dominant contribution to the transverse
response arises from the interference between the OB and MEC, while
the pure MEC contribution is relatively small.  It is also worth
stressing that while the pure MEC contribution is, of course,
positive, the interference contribution is negative.

\begin{figure*}
\centering
\includegraphics[width=7cm,bb=110 270 460 770]{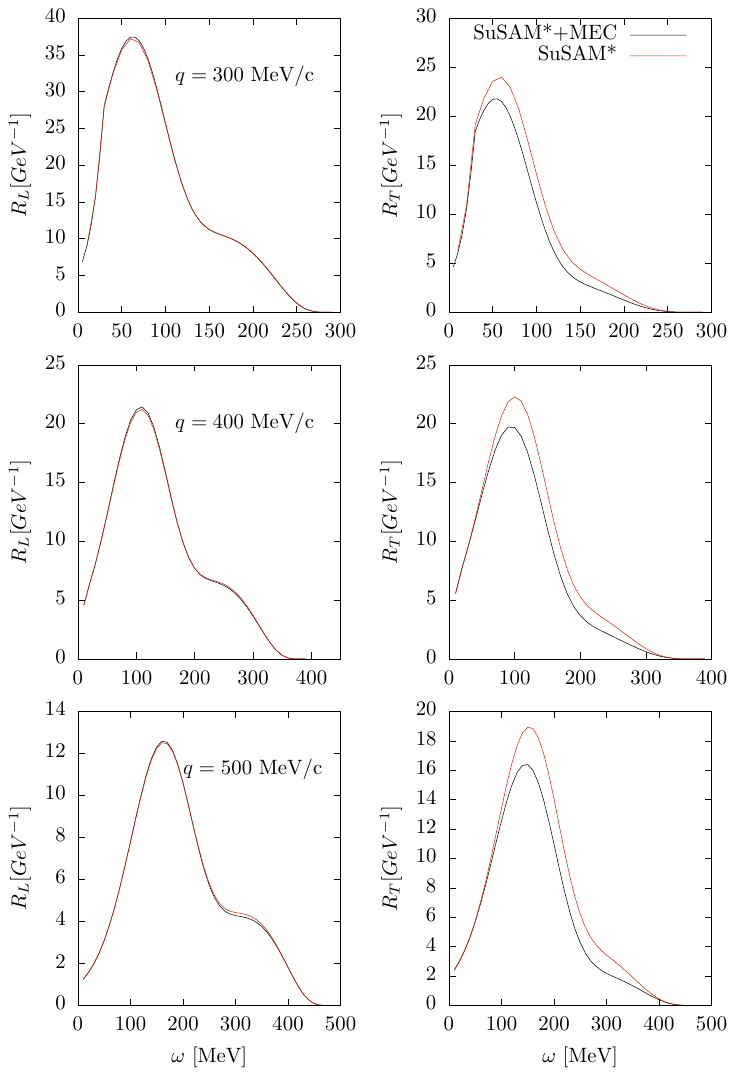}
\kern 4mm
\includegraphics[width=7cm,bb=110 270 460 770]{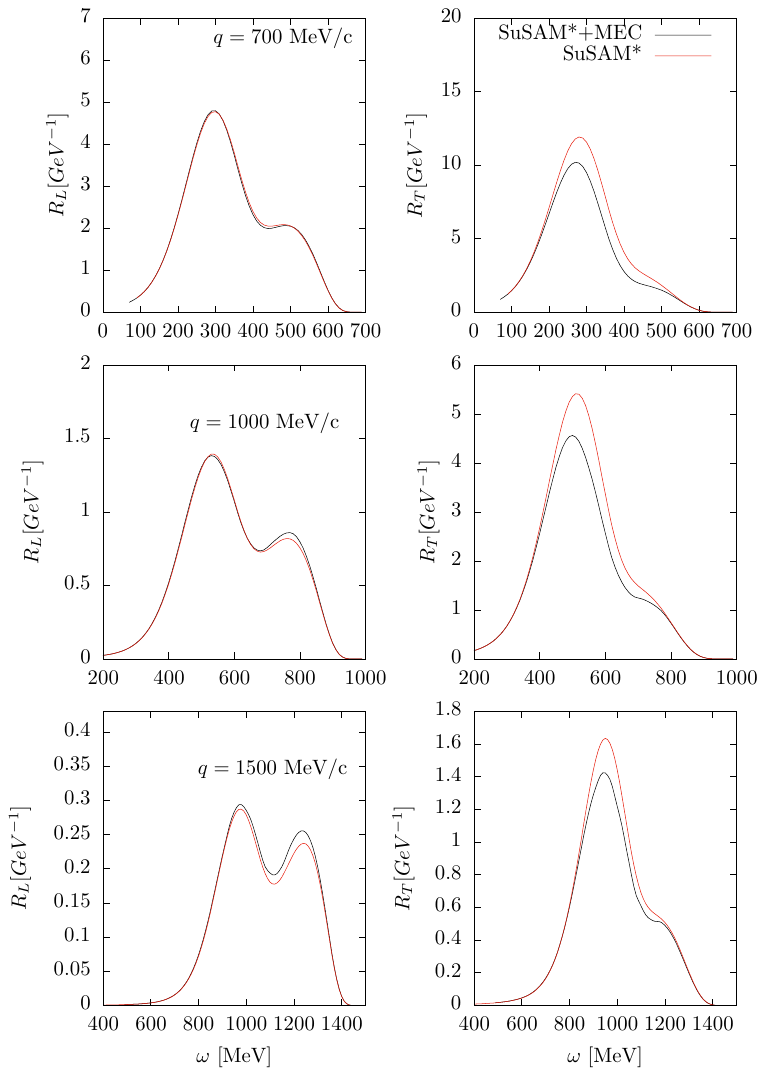}
\caption{
       Response functions calculated in the generalized SuSAM*
       model (black curves). The red curves do not include the MEC. 
 }
 \label{fig-responses}
\end{figure*}

In Fig. \ref{fig-responses}, we present the total responses of
$^{12}$C computed using the generalized SuSAM* model. These responses
are obtained by multiplying the phenomenological scaling function by
the averaged single-nucleon response and summing over protons and
neutrons, as given by Eq. (36). The responses are shown for different
values of $q$ as a function of $\omega$. In the same figure, we also
show the results without including the MEC contributions, which
corresponds to setting the terms $w_{12}+w_{2}$ associated with the
two-body current (Eq. (21)) to zero.

Comparing the results with and without MEC, we observe that the impact
of MEC is more significant in the transverse response compared to the
longitudinal response. This is expected since the corrections due to
MEC in the longitudinal response are higher-order effects in a
non-relativistic expansion in powers of $v/c$, as
known from previous studies \cite{Ris89}. Therefore, the MEC
contributions to the longitudinal response are minimal and only start
to become noticeable for $q>$1 GeV in the high-energy region. However,
this high-energy region is dominated and overshadowed by pion emission
and inelastic processes, making it difficult to isolate
the 1p1h longitudinal response.

The inclusion of MEC in the single-nucleon leads to a reduction of the
transverse response by around 10\% or even more for all studied values
of $q$. This is consistent with previous calculations in RFG and the
shell model \cite{Ama94,Fab97,Ama03,Ama03b,Ama10}. These calculations
have consistently shown that MEC in the 1p1h channel tend to decrease
the transverse response compared to the contribution from the one-body
current.  It is important to note that this reduction in the
transverse response is a direct consequence of the destructive
interference between the one-body current and MEC. The contribution of
MEC to the transverse response is negative because the direct two-body
matrix element is zero (in symmetric nuclear matter, $N=Z$) or almost
zero (in asymmetric nuclear matter, $N\ne Z$, or in finite nuclei)
after summing over isospin.

\begin{figure}
\centering
\includegraphics[width=7cm,bb=125 340 400 770]{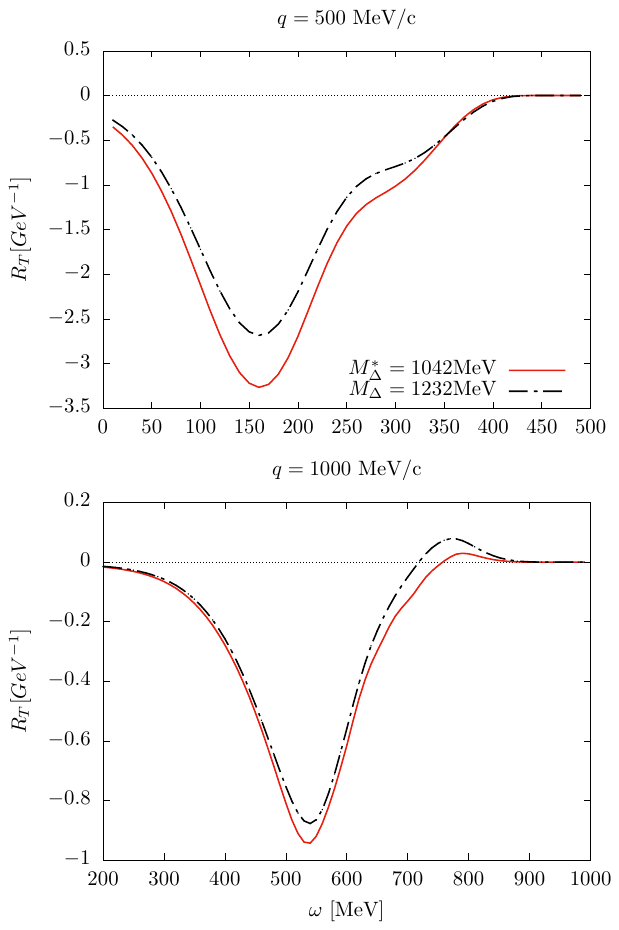}
  \caption{Comparison of the transverse interference OB-$\Delta$ computed 
in the generalized SuSAM* model with and without  
relativistic 
 effective mass and vector energy for the $\Delta$.}
  \label{fig10}
\end{figure}

The treatment of the $\Delta$ resonance in the medium is subject to
various ambiguities and uncertainties. In our generalized SuSAM*
model, we have assumed that the $\Delta$ resonance acquires an effective
mass $M_\Delta^*$ and vector energy $E_v^\Delta$ due to its interaction with the
RMF. This requires modifying the propagator according to the formalism
proposed in references \cite{Weh93,Kim96}.
To estimate the effect of this treatment, in Fig. 14 we compare the
transverse response for the OB-$\Delta$ interference calculated assuming
that the $\Delta$ remains unchanged in the medium, i.e., setting
$M_\Delta^* = M_\Delta$ and $E_v^\Delta = 0$. The response with the
free $\Delta$ without medium modifications is slightly smaller in
absolute value, around 10\% depending on the momentum transfer. This
can be seen as an estimation of the uncertainty associated with the
$\Delta$ interaction in the medium.

Another related issue is the modification of the $\Delta$ width in the
medium, which we have not considered here assuming the free width
\eqref{width}. This effect can also influence the results, but it is
expected to be of the same order as the observed effect in Fig. 14.
It is important to note that the treatment of the $\Delta$ resonance
in the medium is a complex topic, and further investigations and
refinements are needed to fully understand its effects and
uncertainties.

\begin{figure*}
\includegraphics[scale=0.85]{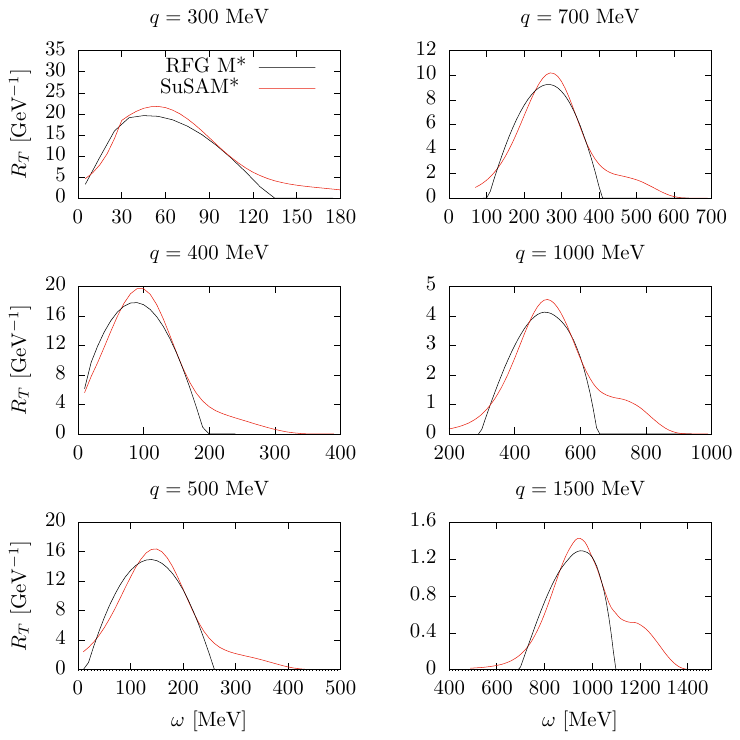}
\vspace*{-14.0cm}
\caption{Total transverse responses for $^{12}$C including MEC in the
  RMF model with $M^*=0.8$ compared to the generalized SuSAM* model.
}
\label{fig-rt}
\end{figure*}

In Fig. \ref{fig-rt}, we compare the total transverse response
calculated in the RMF model with an effective mass of $M^*=0.8$ to the
results obtained in the generalized SuSAM* approach for various
momentum transfers, ranging from $q=300$ MeV/c to $q=1500$ MeV/c. Both
calculations include the effects of MEC.  One notable difference
between the two approaches is the presence of a pronounced tail at
high energy transfer rates in the SuSAM* results. This tail extends
well beyond the upper limit of the RFG responses, reflecting the
effect of the phenomenological scaling function used in the SuSAM*
approach. Similar effects are found in the longitudinal response. 
 Additionally, it is worth noting that the peak height
of the transverse response in the SuSAM* approach is generally higher
compared to the RMF model.  Overall, the comparison in
 Fig. \ref{fig-rt} highlights the improvements and additional physics
captured by the SuSAM* approach, by extending the scaling function of
the RFG to describe the transverse response in a wider energy transfer
range.

\begin{figure}
\centering
\includegraphics[width=8cm,bb=50 275 525 770]{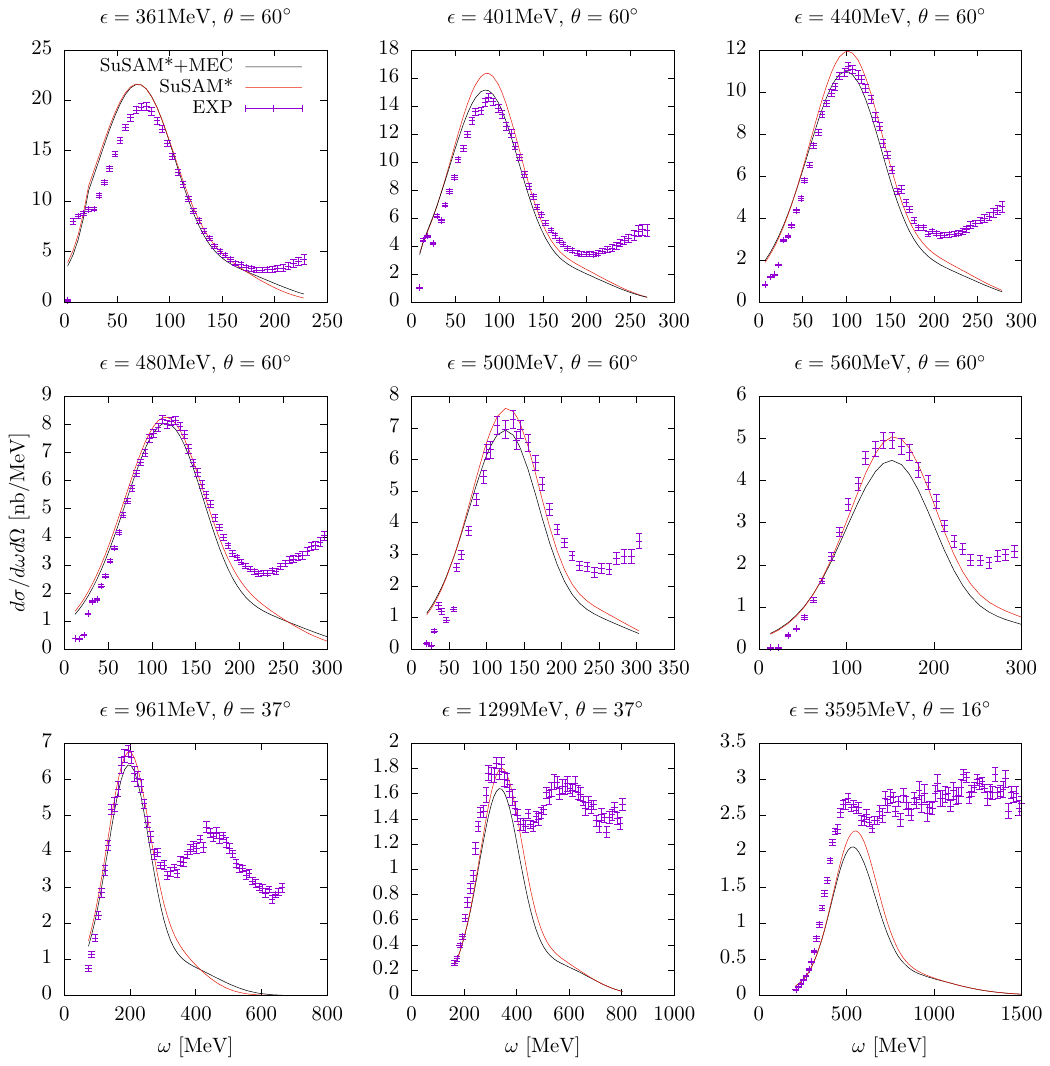}
  \caption{Cross section of $^{12}$C for several kinematics computed
    with the generalized SuSAM* model, including MEC, compared with
    the same calculation without MEC. Experimental data are from
    Refs. \cite{archive,archive2}.}
  \label{fig10}
\end{figure}

Finally, in Fig. 16, we present the results for the (e,e') double
differential cross section of $^{12}$C calculated with the generalized
SuSAM* model including MEC, compared to experimental data for selected
kinematics. We also compare with the same model but assuming that only
the single-nucleon contribution is present, i.e., setting the MEC to
zero. We observe that the inclusion of MEC in this model leads to a
small reduction in the cross section compared to the case without
MEC. This reduction is a consequence of the decrease in the transverse
response due to the presence of MEC.  The generalized scaling
approach, including the inclusion of MEC, provides a global
description of the cross section that is comparable to other previous
analyses, such as the SuSAM* model with the one-body current
 only, or the SuSAv2 model, which factorize
different definitions of the single nucleon (without effective mass
and with extrapolation of the Fermi gas single nucleon in the case of SuSAv2). All of these approaches reasonably
describe the quasielastic cross section because the scaling function
has been properly adjusted to reproduce the global scaling data.  The
generalized scaling approach, like any parametrization, is a
phenomenological framework that aims to capture the essential physics
of the reaction. It provides a functional form for the cross section
that incorporates the known ingredients and leaves the unknowns to be
determined by the scaling function. The scaling function encapsulates
the effects of various dynamical and correlation effects, allowing for
a global description of the data.

\section{Discussion and concluding remarks}

From the results seen in the previous section we have observed that,
in all the models considered, the transverse response decreases when
including meson exchange currents in the 1p1h channel.  This result is
consistent with previous independent calculations performed in the
relativistic and non-relativistic Fermi gas models as well as in the
non-relativistic and semi-relativistic shell models.  The result is a
consequence of the fact that the main contribution arises from the
interference of the OB and $\Delta$ currents, in particular through
the exchange diagram, carrying a minus sign.  The contribution from
the direct part of the MEC matrix element is zero in the Fermi gas,
and this is the reason for the negative contribution.

It is worth mentioning the existence of some calculations
that disagree with this result and suggest a different effect of MEC
on the transverse response.  We would like to comment in particular on
two notable model calculations: the Green Function Monte Carlo (GFMC)
model from reference \cite{Lov16} and the Correlated Basis Function
(CBF) calculation by Fabrocini \cite{Fab97}, both including meson
exchange currents in the 1p1h sector.  In both approaches, the effect
of MEC is positive in the quasielastic peak and quite significant,
around 20\%, in the transverse response. This substantial effect is
attributed to the simultaneous effect of tensor correlations in the
wave function and MEC. In fact, in Fabrocini's calculation, the origin
of this effect was found to be the tensor-isospin correlation
contribution in the direct matrix element of the $\Delta$ current,
which is non-zero when summing over isospin for correlated wave
functions.  This effect can also be understood in terms of presence of
short-range correlations in the nuclear wave function. The direct
matrix element of MEC, when a proton is emitted, involves the
interaction of the proton with protons as well as with neutrons, i.e.,
the MEC matrix element involves PN and PP pairs. The high-momentum
component of these pairs is significantly different because PN pairs
contain the $^3S_1$ and $^3D_1$ deuteron-like waves, while PP pairs do
not. Therefore, when summing over isospin, there is no cancellation
between PP and PN pairs in the high-momentum part of the wave
function, resulting in a non-zero direct matrix element. This is in
agreement with the conclusion of Fabrocini, as the tensor-isospin term
precisely generates this significant difference between PP and PN
pairs.  An alternative way to investigate this hypothesis would be to
perform calculations in the independent particle model by solving the
Bethe-Goldstone equation \cite{Cas23b} for PP and PN pairs and using a
correlation current similar to the one proposed in \cite{Mar23}.  Such
calculations could provide further insights into the effect of
short-range correlations on the MEC contributions to the transverse
response.

On the other hand the results of Fabrocini reproduce the well-known
effect that MEC has a negative impact on the transverse response when
the correlations functions are set to zero, consistent with the
results from uncorrelated models.
Since in the present work we started with an uncorrelated
model, the relativistic mean field, the effects of correlations in the
transverse current are expected to be included phenomenologically in
the scaling function. This and other mechanisms, such as final state
interactions, contribute to the violation of scaling observed in the
data.

To 
 summarize, this work presents a generalized scaling analysis of
the (e,e') cross section of $^{12}$C, including the MEC consistently
in the formalism. To achieve this, we have introduced a new definition
of the single nucleon tensor in the factorization of the model. The
average per particle of the hadronic tensor for 1p1h emission has been
defined by considering the sum of the one-body and two-body currents,
without modifying the definition of the scaling function, which
remains the same as in the one-body current case in the Fermi
gas. This averaging definition has been extended beyond the scaling
region $-1 < \psi^* < 1$ of the Fermi gas by slightly modifying the
momentum distribution with a smeared Fermi distribution that allows
the evaluation of MEC for any value of the scaling variable.

By incorporating the MEC and using the phenomenological scaling
function, we have calculated the 1p1h response functions in the RFG,
RMF, and SuSAM* models. The results show the impact of the MEC on the
response functions, particularly in the transverse sector. The MEC
reduce the transverse response while the longitudinal response is
found to be hardly affected by the MEC.  Furthermore, the analysis of
the OB-MEC interference and the comparison between the SuSAM* and RFG
models highlight the role of the effective mass and the $\Delta$
resonance in the response functions.

Overall, the generalized scaling analysis with the inclusion of MEC
provides a consistent framework for studying quasielastic electron
scattering in nuclei accounting for relativistic dynamical effects
through the effective mass. The approach adopted in this work differs
from other scaling analyses, such as the original SuSAM* model, in the
definition of the single-nucleon dividing factor, which now
incorporated the effect of MEC in the 1p1h channel. However, the
ultimate results are compatible between different models because the
improvement in scaling symmetry is not significant when modifying the
single nucleon in this manner. This means that both formalisms will
describe the experimental cross section data similarly, as they have
been adjusted accordingly. The difference between various approaches
lies in how the scaling function is adapted and rectified based on the
chosen prefactor of the single nucleon. The equivalence between these
models and others, such as SuSAv2, indicates the flexibility of the
scaling approach to adapt to the circumstances of the emphasized
model. Scaling is only an approximate symmetry of quasielastic data,
and the degree of violation of this symmetry should be attributed to
all effects that break the factorization of the cross section in a
many-body system with complex interactions and correlations between
particles.

In conclusion, this work presents the first comprehensive study of
quasielastic electron scattering in nuclei that includes the 1p1h meson
exchange currents (MEC) consistently in a generalized scaling approach,
extending previous work where this contribution was evaluated in the
relativistic Fermi gas (RFG) framework.
Looking ahead, this work opens the door to future
developments and applications, including the extension of the model to
study neutrino-nucleus scattering.

\section{Acknowledgments}

Work supported by:
Grant PID2020-114767GB-I00
funded by MCIN/AEI/10.13039/501100011033; FEDER/Junta de
Andalucia-Consejeria de Transformacion Economica, Industria,
Conocimiento y Universidades/A-FQM-390-UGR20; Junta de
Andalucia (Grant No. FQM-225); 
INFN under Project  NUCSYS; and
University of Turin under Project BARM-RILO-23-01.


\end{document}